\documentclass[apjl,twocolumn]{emulateapj_mod}
\usepackage{epsfig,apjfonts,mathptmx}

\def\gtsima{$\; \buildrel > \over \sim \;$}
\def\ltsima{$\; \buildrel < \over \sim \;$}
\def\prosima{$\; \buildrel \propto \over \sim \;$}
\def\gsim{\lower.5ex\hbox{\gtsima}}
\def\lsim{\lower.5ex\hbox{\ltsima}}
\def\simgt{\lower.5ex\hbox{\gtsima}}
\def\simlt{\lower.5ex\hbox{\ltsima}}
\def\simpr{\lower.5ex\hbox{\prosima}}

\def\h1{$h^{-1}$}
\def\eeq{\end{equation}}
\def\beq{\begin{equation}}
\def\24mu{24\,$\mu{\rm m}$}
\def\70mu{70\,$\mu{\rm m}$}
\def\8mu{8\,$\mu{\rm m}$}

\shorttitle{$z=2$ star-formation in GOODS}
\shortauthors{E. Daddi et al.}

\begin{document}

\title{Multiwavelength study of massive galaxies at
\lowercase{z}~$\sim2$.\\ I.  Star formation and galaxy growth}

\author{
E. Daddi\altaffilmark{1},
M. Dickinson\altaffilmark{2},
G. Morrison\altaffilmark{3,4},
R. Chary\altaffilmark{5},
A. Cimatti\altaffilmark{6},
D. Elbaz\altaffilmark{1},
D. Frayer\altaffilmark{5},
A. Renzini\altaffilmark{7},
A. Pope\altaffilmark{8},
D.M. Alexander\altaffilmark{9},
F.E. Bauer\altaffilmark{10},
M. Giavalisco\altaffilmark{11},
M. Huynh\altaffilmark{5},
J. Kurk\altaffilmark{12},
M. Mignoli\altaffilmark{13}
}

\altaffiltext{1}{Laboratoire AIM, CEA/DSM - CNRS - Universit\`e Paris Diderot,
DAPNIA/SAp, Orme des Merisiers,  91191 Gif-sur-Yvette, France {\em edaddi@cea.fr}}
\altaffiltext{2}{National Optical Astronomy Observatory,  950 N. Cherry
Ave., Tucson, AZ, 85719}
\altaffiltext{3}{Institute for Astronomy, University of Hawaii, Honolulu,
HI, 96822, USA}
\altaffiltext{4}{Canada-France-Hawaii Telescope, Kamuela, HI, 96743, USA}
\altaffiltext{5}{{\em Spitzer} Science Center, Caltech, MS 220-6, CA 91125}
\altaffiltext{6}{Dipartimento di Astronomia, Universit\'a di Bologna, Via
Ranzani 1, I-40127, Bologna, Italy}
\altaffiltext{7}{INAF, Osservatorio Astronomico di Padova, Vicolo Osservatorio 5, I-35122, Padova, Italy}
\altaffiltext{8}{Department of Physics \& Astronomy, University of British Columbia, Vancouver, BC, V6T 1Z1, Canada}
\altaffiltext{9}{Department of Physics, Durham University, South Road, Durham, DH1 3LE, UK}
\altaffiltext{10}{Columbia Astrophysics Laboratory, Columbia University, Pupin Laboratories, 550 West 120th Street, Roomm 1418, New York, NY 10027}
\altaffiltext{11}{University of Massachussets, Astronomy Department,
Amherst, MA 01003, USA}
\altaffiltext{12}{Max-Planck-Institut f\"ur Astronomie, K\"onigstuhl 17,
D-69117, Heidelberg}
\altaffiltext{13}{INAF, Osservatorio Astronomico di Bologna, via Ranzani, 1 - 40127 Bologna, Italy}

\begin{abstract}
Examining a sample of massive galaxies at $1.4<z<2.5$ with $K_{Vega}<22$
 from the Great Observatories
Origins Deep Survey, we compare photometry from Spitzer at mid- and far-IR, to
submillimeter, 
radio and rest-frame ultraviolet wavelengths, to test the agreement between 
different tracers of star formation rates (SFRs) and to explore the implications 
for galaxy assembly.  For $z\sim2$ galaxies with moderate luminosities 
($L_{8\mu m}<10^{11}L_\odot$), we find that the SFR can be estimated consistently 
from the multiwavelength data based on local luminosity correlations.  However, 
20--30\% of massive galaxies, and nearly all those with $L_{8\mu m}>10^{11}L_\odot$,
show a mid-IR excess which is likely due to the presence of obscured active
nuclei, as shown in a companion paper.  There is a tight and roughly linear correlation
between stellar mass and SFR for \24mu-detected galaxies.  For a given mass, the SFR 
at $z=2$ was larger by a factor of $\sim 4$ and $\sim 30$ relative to that in 
star forming galaxies at $z=1$ and $z=0$, respectively.  Typical ultraluminous 
infrared galaxies (ULIRGs) at $z=2$ are relatively 'transparent' to ultraviolet 
light, and their activity is long lived ($\simgt400$~Myr), unlike that in local 
ULIRGs and high redshift submillimeter-selected galaxies.  ULIRGs are the common 
mode of star formation in massive galaxies at $z=2$, and the high duty cycle 
suggests that major mergers are not the dominant trigger for this activity.  
Current galaxy formation models underpredict the normalization of the mass-SFR 
correlation by about a factor of 4, and the space density of ULIRGs by an order 
of magnitude, but give better agreement for $z>1.4$ quiescent galaxies.
\end{abstract}

\section{Introduction}

Tracing the buildup of galaxies at high redshifts requires reliable
measurements of two fundamental quantities: stellar masses and
star-formation rates (SFR).  The integrated history of star-formation rates
within galaxies provides a lower limit to their stellar mass growth.
Stellar mass growth in excess of this contribution is to be ascribed
to galaxy mergers. Most of the rise in the space density of massive
galaxies takes place in the cosmic time interval between $z\sim 3$ and
$z\sim 1$ (e.g., Dickinson et al. 2003; Rudnick et al. 2003; 2006; 
Fontana et al.\ 2004; 2006; Daddi et al.\ 2004a, D04a hereafter; van Dokkum
2006; Franceschini et al. 2006; Pozzetti et al.\ 2007; Arnouts et al. 2007).  It is still unclear, however, what is the
major channel for galaxy growth during this period, and the
quantitative role of mergers in building up the mass of
galaxies. Violent starbursts have been found to be common in the
$z\sim2$ universe (D04a, Daddi et al. 2005b, D05b hereafter; Shapley et al.\
2005; F\"orster-Schreiber et al.\ 2004a), but their relevance on the
mass growth of galaxies critically depends on the effective duration
of these vigorous star-formation events (see D05b). To
make substantial progress one needs to look at the distribution of
star-formation rates in distant galaxies, in order to ascertain
the relative frequency of these events among massive galaxies. This
requires accurate measurements of individual star-formation rates in
large samples of distant galaxies.

The advent of the Spitzer Space Telescope holds substantial promise to
achieve these goals, as it now routinely allows one to detect distant
galaxies with the Multiband Imaging Photometer for Spitzer (MIPS) at \24mu 
all the way to redshift of $\sim 3$ (and even beyond). This corresponds 
to the rest-frame spectral mid-IR
region (5--12$\mu$m), which tends to be dominated by polycyclic
aromatic hydrocarbon features (PAH).  In the local universe, the
mid-IR luminosities of galaxies are known to correlate with the total
infrared luminosities ($L_{\rm IR}$; see Chary \& Elbaz 2001; Elbaz et al.
2002; Dale \& Helou 2002; F\"orster-Schreiber et al.\ 2004b), a well
established probe of the star-formation rate of galaxies. Problems
however exist because the PAH emission is known to be reduced at low
metallicities (Engelbracht et al.\ 2005; Madden et al.\  2006). Spitzer
observations show a substantial scatter in the local correlation of
mid-IR luminosity vs. the total infrared luminosity $L_{\rm IR}$ (Dale et
al.\ 2005; Smith et al.\ 2006; Armus et al.\ 2006).  In addition, a
substantial contribution to the mid-IR light can come from the
accretion disks of Active Galactic Nuclei (AGNs). AGNs in distant
galaxies can be identified by their X-ray emission, but this would not be
detectable at $z=2$ in the case of large column densities of obscuring
material even for luminous AGN (e.g., as in the case of NGC~1068).

Several recent surveys have used the \24mu    emission in $1<z<3$
galaxies in order to estimate their ongoing star-formation rates and
related quantities (e.g., P\'erez-Gonz\'alez et al.\ 2005; Reddy et al.\
2005; D05b, Papovich et al.\ 2006; 
Caputi et al.\ 2006ab; etc.).  This is
generally accomplished by using prescriptions relating the mid-IR
luminosity to $L_{\rm IR}$, as calibrated in the spectral libraries from
Chary \& Elbaz (2001; CE01 hereafter) and Dale \& Helou (2002; DH02
hereafter), although sometimes other empirical conversions have been
used (see, e.g., Caputi et al.\ 2007).  However, it remains to be established
to which extent these estimates are still accurate for $z\sim2$
galaxies. For example, the impact of the silicate 9.7\,$\mu$m
absorption feature at high redshift has still to be quantified, and
the relative strength of the PAH emission as a function of star
formation activity might be different at high redshift with respect
to the local Universe.

Using multi-wavelength data from the Northern field of the Great
Observatories Origins Deep Survey (GOODS), D05b showed
that the average SED of massive $z=2$ galaxies is consistent with that
of present--day ULIRGs. This implies, on average, a good agreement between
star-formation tracers as calibrated locally and based on the radio,
far-IR, mid-IR and even UV (corrected for the extinction by dust) and
the X-ray. This agreement is encouraging in view of our need to
estimate the star-formation rates in distant galaxies, especially given
the currently limitations in probing the luminosities of galaxies at
$z\sim2$ at the peak of the far-IR emission at $\sim60-100\;\mu$m 
in the rest-frame. 
In this regard, the situation will eventually improve with Herschel,
and later on with ALMA.  In D05b a number of
interesting features of high redshifts ULIRGs were discussed, e.g.,
that they may have fairly long duty cycles and that they appear to be
relatively 'transparent' to the UV photons. These and other aspects are
visited again here in more detail.

The main aim of this paper is to verify to which extent this agreement
among the various star formation tracers persists individually for
distant massive galaxies, hence to which extent SFRs can be
unambiguously measured at high redshift. In particular, reliable SFRs
are instrumental for relating to them other galaxy properties in a
meaningful way. For example, recent surveys have shown that a tight
correlation exists at $z=0$ and at $z=1$ between the stellar mass and
the SFRs of blue, actively star forming galaxies (Noeske
et al.\ 2007; Elbaz et al.\ 2007). It is not yet clear if such a
fundamental correlation exists also at higher redshifts, when most of
the stars in massive galaxies were formed.

For this work we make use of some of the deepest multiwavelength
datasets, available as part of GOODS. 
In particular, compared to D05b, we use here 
a much deeper 1.4~GHz dataset over GOODS-North (GOODS-N hereafter) obtained with the Very 
Large Array (VLA) (Morrison et al.\ in preparation), and \70mu photometry 
from both GOODS fields from Frayer et al.\ (2006).  Also crucial to this work 
is the large dataset of spectroscopic redshifts that is available to us, 
including the VLT/FORS2 spectroscopic redshifts for the GOODS-South field 
(GOODS-S hereafter; Vanzella et al.\  2005, 2006) and those from the ultradeep spectroscopy of
the GMASS survey (Kurk et al., in preparation).  Finally, we have used
in this work the latest realization of galaxy formation models based
on the Millennium simulations (Springel et al.\ 2005), in the form of
mock lightcones presented by Kitzbichler \& White (2007). This allows
us to constrain the models and gain insights into the nature of star
forming galaxies at high redshifts.

The paper is organized as follows. The sample selection of $z\sim2$
galaxies, with spectroscopic and photometric redshifts are presented
in Section~\ref{sec:data}. The multiwavelength datasets used are presented
in Section~\ref{sec:multi}, with a discussion of known limitations of
each dataset as a star formation indicator, especially for high
redshift galaxies. Section~\ref{sec:mips} compares the \8mu rest-frame   
luminosities of galaxies
(estimated from their \24mu flux densities) 
to those at other wavelengths, while
Section~\ref{sec:uv} focuses on the use of ultraviolet (UV) luminosities
and on the reliability of extinction corrections. A suggested recipe for
estimating SFRs  in $z=2$ galaxy samples is given in
Section~\ref{sec:recipe}.  We present the implications of our findings
for the characterization of star formation in distant galaxies, 
galaxy growth and the
comparison with theoretical models in Section~\ref{sec:impli}. Our
conclusions are summarized in Section~\ref{sec:end}.  We assume a
Salpeter initial mass function (IMF) from 0.1 and 100 $M_\odot$, and a
WMAP3 cosmology with $\Omega_\Lambda, \Omega_M = 0.76, 0.24$, and $h =
H_0$[km s$^{-1}$ Mpc$^{-1}$]$/100=0.73$ (Spergel et al.\ 2007).
Throughout the paper we use magnitudes in the Vega scale, unless stated
otherwise.

\section{Samples of $z\sim2$ galaxies in GOODS}
\label{sec:data}

The sample of galaxies analyzed in this study has been selected in the
$K$-band, and applying the $BzK$ selection technique introduced by
Daddi et al. (2004b; D04b hereafter), that allows us to define highly complete samples
of galaxies in the redshift range $1.4<z<2.5$.  Unless explicitly
stated otherwise, we limit the analysis to $BzK$ star forming galaxies
(or $sBzK$s, as opposed to the passive ones or $pBzK$s), as we aim
to focus on galaxies with active star formation.  In the GOODS-N field
we have selected 273 galaxies to $K<20.5$, corresponding
to the 5-sigma detection limit of the data taken with the Flamingos
camera on the Mayall 4m NOAO telescope (the dataset used here is
slightly deeper than that used in D05b).  In the GOODS-S
field we selected 1018 sources detected with $S/N>5$ down to $K<22$.
The GOODS-S field $K$-band VLT ISAAC data is not
homogeneously deep over the whole field, but reasonably complete over
the whole used area at these $K$-band limits.  We have excluded from
the analysis all hard X-ray detected galaxies in both GOODS fields,
using the catalogs of Alexander et al.\ (2003), and galaxies with power
law SEDs over the 3.6--24 $\mu$m wavelength range observed by the 
Spitzer Infrared Array Camera (IRAC) as likely AGN-dominated
sources. Only a handful of power-law galaxies remain undetected in our
deep X-ray imaging data.  Also, in both fields we only consider
galaxies whose IRAC counterpart is closer than 0.5$''$ from the K-band
position.  We have empirically verified that this criterion is
efficient at excluding sources where the IRAC bands photometry is
substantially contaminated by blending due to the relatively poor IRAC
PSF of about 1.6$''$. Statistically, this excludes some 10\% of the
galaxies, for the accurate IRAC detection procedures carried out in
GOODS and based on mexhat kernel detections using SExtractor
(Dickinson et al., in preparation). This criterion is required in order
to have reliable photometry over the IRAC bands (hence solid
photometric redshift estimates), and to avoid problems with
blending in the MIPS \24mu    photometry. The  \24mu    flux densities (Chary et
al., in preparation) are in fact estimated using prior source positions
from the IRAC catalogs.

\begin{figure*}
\centering
\includegraphics[width=18cm]{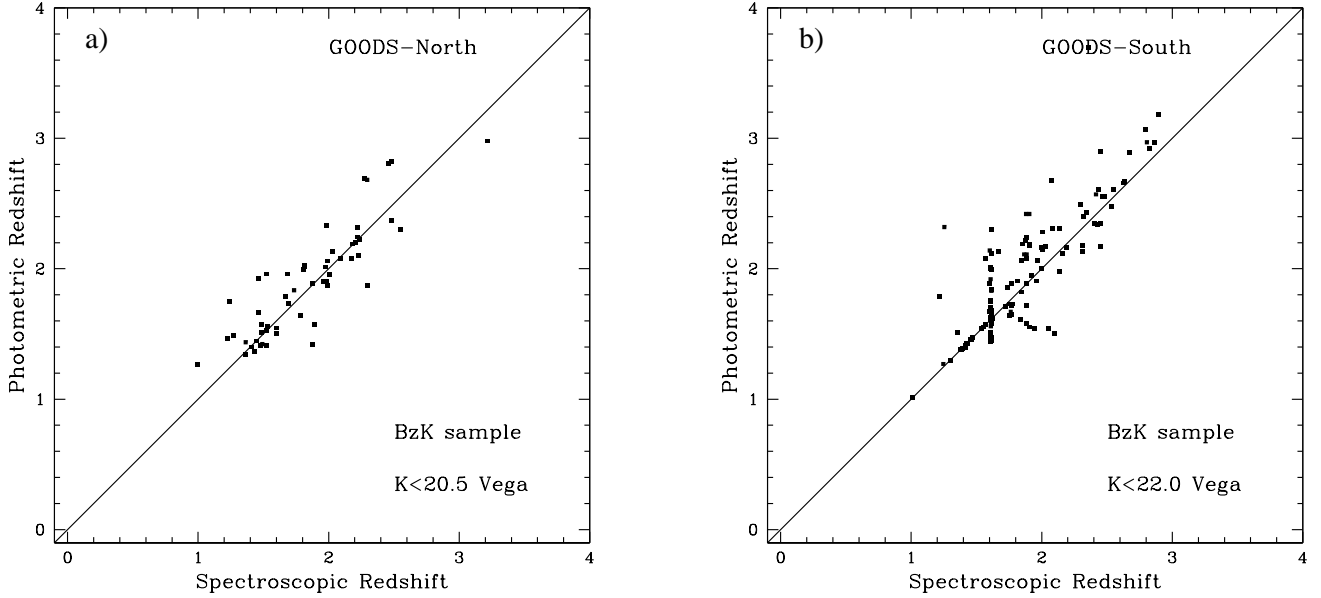}
\caption{Comparison of spectroscopic and photometric redshift for galaxies
in GOODS-N (left panel) and in GOODS-S (right panel). The GOODS-S photometric
redshifts that we use are taken from Grazian et al.\ (2006), and we show in
the right panel of 
this figure only those galaxies that did not already have a spectroscopic 
redshift in Grazian et al.\ (2006).
}
\label{fig:zphot_N}
\end{figure*}

\begin{figure*}
\centering
\includegraphics[width=18cm]{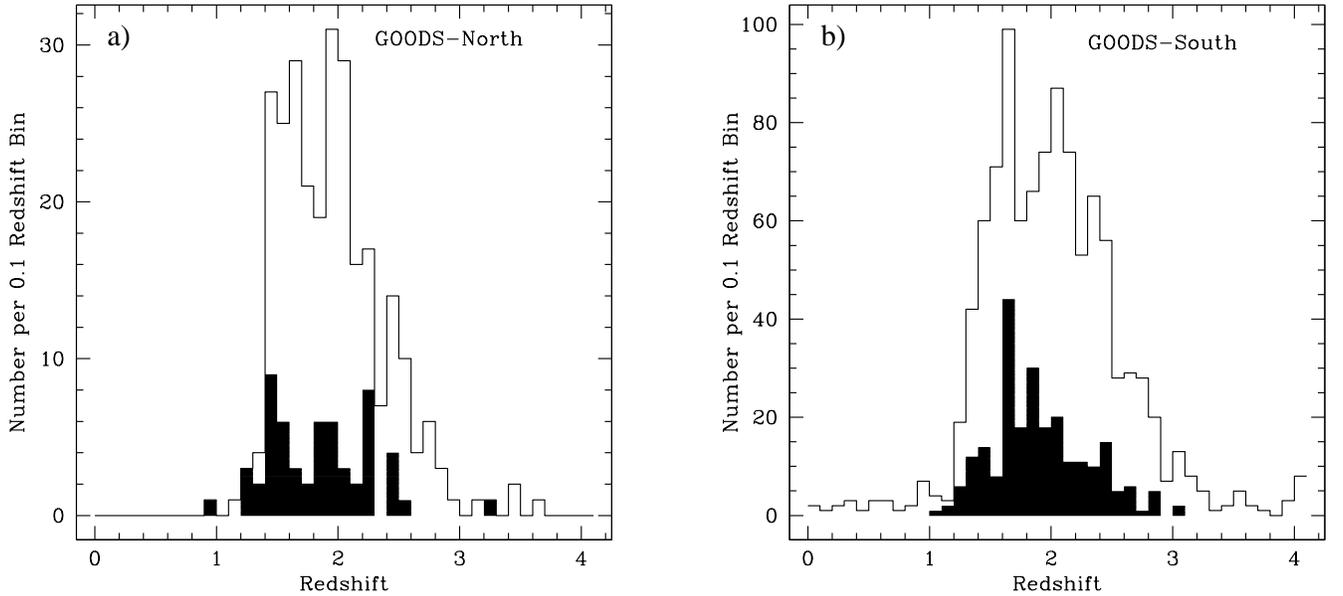}
\caption{The redshift distribution of the $z\sim2$ sample of galaxies
considered in this paper for GOODS-N (left panel) and GOODS-S (right panel).
Solid histogram refer to spectroscopic
redshifts, while the empty histogram is for photometric redshifts for $BzK$
selected star-forming galaxies.
}
\label{fig:zhist_N}
\end{figure*}

In both fields we used multiband photometry in the optical and near-IR
(see Giavalisco et al.\ 2004 for datasets descriptions), updated to the
latest releases and observations. SEDs were derived using the method
described, e.g., in D04b and D05b. Photometric redshifts
were determined in both fields for all $BzK$ selected $z\sim2$ galaxies
by fitting to the empirical templates of Coleman et al.\ (1980), as
described in D04ab. However, for the GOODS-S field we
find that the photometric redshifts measured by Grazian et al.\  (2006),
which include deep U-band data similarly to what we have in GOODS-N,
are in slightly better agreement with the spectroscopic redshifts and
we decided to adopt them for the present analysis.

Spectroscopic redshifts are available in GOODS for a fairly
representative fraction of the galaxy sample considered in this
paper. When limiting to spectroscopic redshifts with flags warranting
reliability, redshifts are available for 175 galaxies in GOODS-S (17\%
of the total sample) and $57$ in GOODS-N (21\% of the sample). For
GOODS-S, in addition to all sources of publicly available redshifts
(VVDS: Le Fevre et al.\ 2004; GOODS/FORS2: Vanzella et al.\ 2005, 2006;
K20 survey: Mignoli et al.\ 2005; CDFS survey: Szokoly et al.\ 2004) we
have most notably used results of redshift surveys developed as part
of GOODS or in collaboration with GOODS, that explicitly targeted the
$z\sim2$ galaxies. This includes the Galaxy Mass Assembly
Spectroscopic Survey (Kurk et al.\ 2006, and in
preparation\footnote{see
http://www.arcetri.astro.it/cimatti/gmass/gmass.html}) that targeted
$z_{phot}>1.4$ galaxies selected at 4.5$\mu$m (IRAC channel 2) to a
limiting AB magnitude = 23.
For the GOODS-N field major sources of $z>1.4$ spectroscopic redshifts
are
Keck LR-blue and DEIMOS observations carried out within GOODS (Stern et al.,
in preparation) and the BM/BX samples of Reddy et al.\ (2006b),
in addition to the TKRS survey (Wirth et al.\ 2004).

Fig.~\ref{fig:zphot_N} panels show the comparison of photometric and
spectroscopic redshifts for our samples of $BzK$ selected galaxies.
The dispersion of the residual in the difference between spectroscopic
and photometric redshifts is $\sigma(z_{spec}-z_{phot})\sim0.25$ for
GOODS-N and GOODS-S.  The larger number of outliers in the full
samples, with respect to the $BzK$ samples, shows that the use of the
$BzK$ selection does help in defining a clean sample of $z\sim2$
galaxies, which is substantially free from lower redshift
contaminants.  We have identified a small sample of galaxies in both
fields with  strong discrepancies between spectroscopic and photometric
redshifts. From a careful analysis of the multiwavelength
SED of these outliers, we found that in most cases the spectroscopic redshift
appears to be the wrong one, and we have adopted the
photometric redshift for the analysis.  This does not affect any of
our major conclusions.

Fig.~\ref{fig:zhist_N} panels show the redshift distribution
of the sources used for the analysis of $z\sim2$ SFR indicators in this
paper.
We do not include in the analysis carried out in the next sections the
objects
in the tails of the distributions, with either $z<1.2$ or $z>3$.

\begin{figure}
\centering
\includegraphics[width=8.8cm]{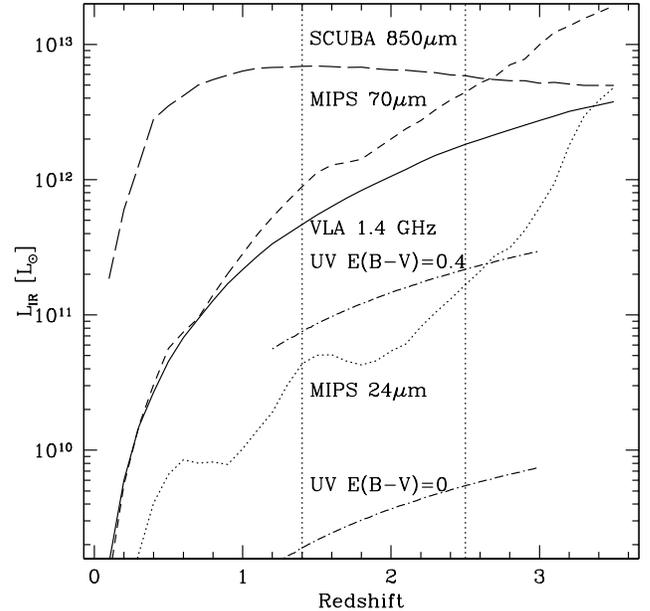}
\caption{The limiting $L_{\rm IR}$ that can be detected with the various
tracers used in this work. We consider limits of 14.1\,$\mu$Jy
(3$\sigma$) for radio 1.4\,GHz, 12\,$\mu$Jy (3$\sigma$) for MIPS \24mu,
1.8\,mJy for MIPS \70mu    (3$\sigma$) and 4\,mJy for 850\,$\mu$m (flux
density of
typical detections, Pope et al.\ 2006, some small regions in GOODS-N
have actually somewhat fainter detections). We use the model of CE01
to convert flux densities into $L_{\rm IR}$ as a function of redshift, except for
radio where we use the radio-IR correlation.  For the UV, the
effective SFR limit in our samples
(formally converted to $L_{\rm IR}$ using Eq.~1) 
is given in principle by the depth of
the B-band imaging ($B=27$ AB, 3$\sigma$), 
from which we derive the UV luminosity. In case of
non-zero dust reddening the reached SFR limit is less deep.  MIPS
\24mu    is by far the deepest long wavelength probe of SFR and IR
luminosities at these high redshifts.  }
\label{fig:pl_limi}
\end{figure}

\section{Multiwavelength datasets, their
interpretation and known limitations}
\label{sec:multi}

A major focus of this paper is the comparison between luminosities at
various wavelengths, each of which is known to correlate with the SFR
in galaxies.  In order to put emission at different wavelengths onto a
common footing for intercomparison, we will adopt nominal conversions
between these observed luminosities and fiducial quantities such as
the total infrared luminosity ($L_{\rm IR}$, integrated from 8 to
1000$\mu$m) or an equivalent star formation rate (SFR).  The use of
$L_{\rm IR}$ as a star formation indicator implies that all energy from
star formation is absorbed by dust and re-radiated in the thermal
infrared, and that there are no other significant sources for dust
heating.  Neither assumption is universally correct: on the one hand
not all UV photons from young massive stars may be absorbed by dust,
whereas emission from an AGN can boost emission at various
wavelengths, including the infrared luminosity $L_{\rm IR}$.  Thus, SFR
estimates based only on $L_{\rm IR}$ would result in underestimates in the
former case, and in overestimates of the true SFR in the latter.
Starburst galaxies with the high SFRs (many tens to hundreds of
$M_\odot$~yr$^{-1}$) and infrared luminosities ($L_{\rm IR} > 10^{11} L_\odot$
are called luminous infrared galaxies or LIRGs; then, for $L_{\rm IR} >
10^{12} L_\odot$ one refers to ultraluminous infrared galaxies, or
ULIRGs; we will also refer to objects with $L_{\rm IR} > 10^{13} L_\odot$
as hyperluminous, or HyLIRGs) characteristic of some the $z \sim 2$
galaxies studied here.
Values for $L_{\rm IR}$ and the SFR may differ substantially, depending on
the specific tracer used to evaluate them. In this section, we present
the different indicators, and give a description of the techniques
used to analyze the data, and of the recipes used to convert the
observed emission to SFRs.  These recipes
are generally based on empirical calibrations in the local universe.
In the next section, we will compare the derived equivalent $L_{\rm IR}$
or SFR values, looking for agreement, which would suggest coherent
behavior from low to high redshift, or for discrepancies, which may suggest
that other effects are coming into play.

Following Kennicutt (1998), we may express the conversion between
$L_{\rm IR}$ and SFR as:

\begin{figure}
\centering
\includegraphics[width=8.8cm]{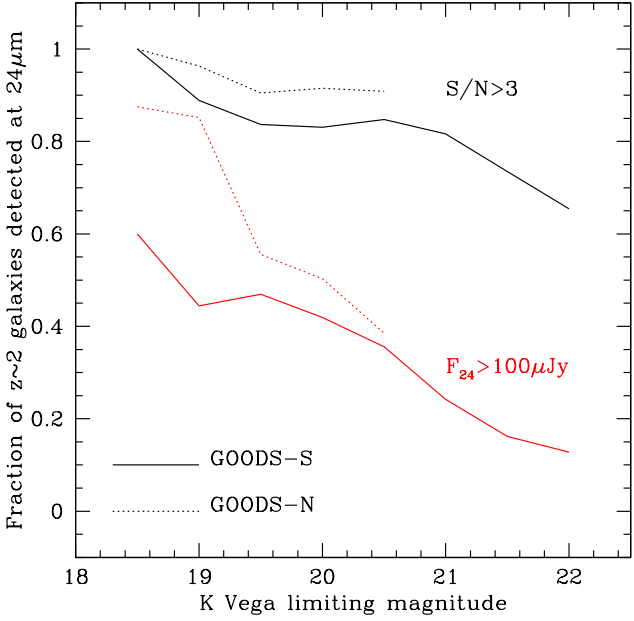}
\caption{The detection rate at \24mu    of the $z\sim2$ galaxies in our
sample
as a function of the K-band limiting
magnitude (cumulative fractions),
for the GOODS-S (solid line) and GOODS-N (dotted line) fields.
Red lines are for a detection limit of 100\,$\mu$Jy, matched by most
shallower surveys performed at \24mu    with Spitzer. The very large
detection
fraction in GOODS up to $K=22$ is in line with the expectations
based on Fig.~\ref{fig:pl_limi}.
}
\label{fig:Det_Rate}
\end{figure}

\beq
{\rm SFR}_{\rm IR} [M_\odot yr^{-1}] = 1.73\times10^{-10} L_{\rm IR} [L_\odot].
\label{eq:LIR}
\eeq This conversion depends on the adopted IMF; here we have adopted
the IMF of Salpeter (1955). Given this relation, SFR$_{\rm IR}$ and
$L_{\rm IR}$ can be seen as equivalent quantities for star forming
galaxies.  In order to estimate the total SFR, we must also include
the contribution of energetic emission from young, massive stars that
is not absorbed and reprocessed by dust.  We estimate it from the
rest-frame UV luminosity as in Eq.~5 of
D04b, with no reddening correction, so to have:

\beq
{\rm SFR}_{\rm tot} = {\rm SFR}_{\rm IR} + {\rm SFR}_{\rm UV,uncorrected}.
\label{eq:SFR}
\eeq

We shall return to the conversion of UV light to SFR in Section~3.6.
The second term above is negligible for red galaxies, as for most of the
$BzK$-selected $z \sim 2$ sources with $K \simlt 20$, but becomes more
relevant at fainter magnitudes when bluer galaxies are more prevalent.

\begin{figure*}
\centering
\includegraphics[width=18cm]{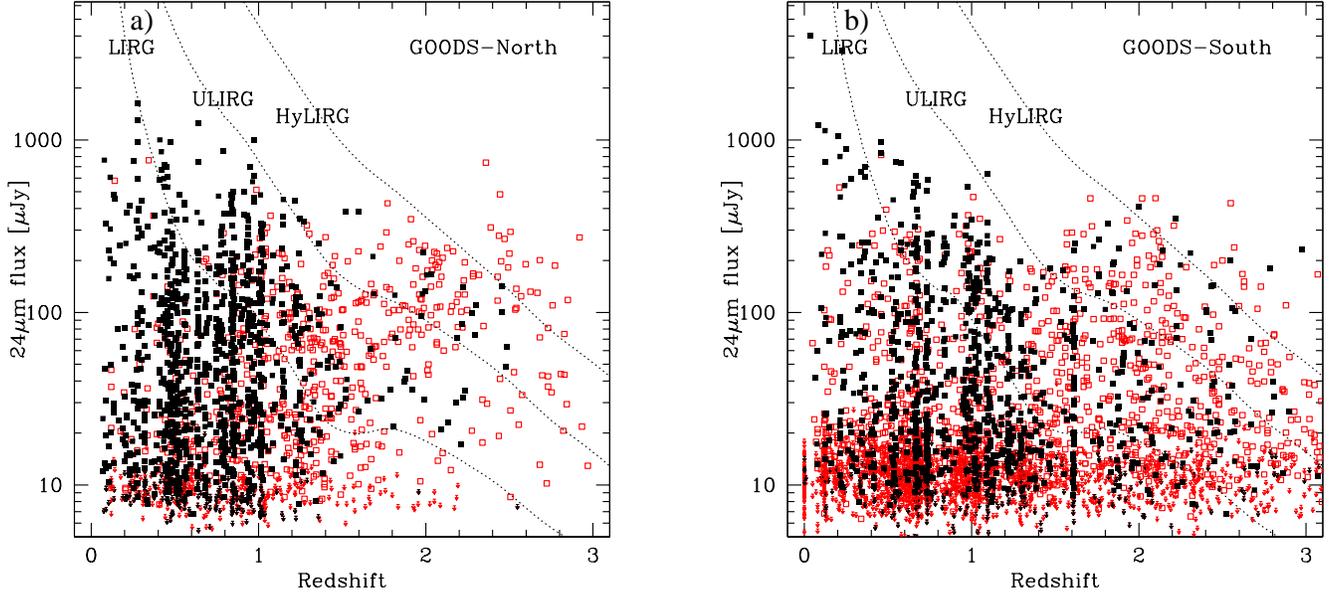}
\caption{Spitzer MIPS \24mu    flux density for galaxies with $K<20.5$ in the
GOODS-N field (left panel) and those with  $K<22.0$ in the GOODS-S field
(right panel).
Black filled symbols are for galaxies with known spectroscopic redshift,
while
red empty symbols are photometric redshifts.
The tracks show expected \24mu    flux densities for galaxies with total IR
luminosity
of $10^{11}, 10^{12}, 10^{13} L_\odot$ (LIRGs, ULIRGs, HyLIRGs), based on the
model SEDs of Chary and Elbaz (2001).
}
\label{fig:24umLIR_N}
\end{figure*}

\subsection{Spitzer MIPS \24mu    imaging}

Both GOODS-N and S have been observed with Spitzer MIPS at \24mu    for about
10 hours per sky pixels.  See D05b for a slightly more
detailed description of these datasets and for the associations of
\24mu    flux densities to K-selected galaxies. Catalogs for the \24mu   
sources will be presented by Chary et al.\ (in preparation). For a
$z=2$ galaxy the \24mu    emission corresponds to approximately an 8\,$\mu$m
rest-frame measurement. For each galaxy in the $1.4\simlt z\simlt 2.5$
redshift range we have derived bolometric infrared luminosities from
the observed flux density at \24mu    and using the luminosity dependent SED
libraries of Chary \& Elbaz (2001) and Dale \& Helou (2002). In
practice, this is done by interpolating $L_{\rm IR}$ over the template
SEDs, sorting the value of $L_{\rm IR}$ that corresponds to the observed
\24mu    flux density.  This determination of the bolometric IR luminosity
implies a large extrapolation, as the $\sim 8\;\mu$m rest-frame
luminosity is generally a small fraction of $L_{\rm IR}$, and therefore
the result is template-dependent. For this reason we also
derived more direct 8\,$\mu$m rest-frame luminosities from the \24mu   
flux density, using the model SEDs (from CE01 and DH02) only to K-correct the
measurement from \24mu    observed to 8\,$\mu$m rest-frame luminosities
(this K-correction is also template dependent but generally small
in the probed redshift range,
within a factor of 2).  For the CE01 templates the relations between
$L(8\,\mu{\rm m})$ and $L_{\rm IR}$ is:

\beq
Log(\frac{L_{\rm IR}}{L_\odot}) = 1.50\times Log(\frac{L_{8\mu m}}{L_\odot}) -
4.31\ \eeq
for $Log(L_{8\mu m}/L_\odot)>9.75$, and

\beq
Log(\frac{L_{\rm IR}}{L_\odot}) = 0.93\times Log(\frac{L_{8\mu m}}{L_\odot}) +
1.23\ \eeq
for $Log(L_{8\mu m}/L_\odot)<9.75$;
\noindent
while for the DH02 templates it is:

\beq
Log(\frac{L_{\rm IR}}{L_\odot}) = 1.21\times Log(\frac{L_{8\mu m}}{L_\odot}) -
1.25.
\label{eq:L8}
\eeq

Here, and in the remainder of the paper, $L(8\mu m)$ is $\nu L_\nu$ at
$8\mu m$.  The two relations give the same
$L_{\rm IR}\sim5\times10^{11}L_\odot$ for $L_{8\mu
m}\sim5\times10^{10}L_\odot$, For higher $L_{8\mu m}$ the CE01
templates result in increasingly larger values of $L_{\rm IR}$, up to a
factor of 2.3 for $L_{8\mu m}=10^{12}L_\odot$. At $L_{8\mu
m}<5\times10^{11}L_\odot$ the CE01 templates results in lower $L_{\rm IR}$
values with respect to DH02. For the range of interest in this work,
the differences are always well within a factor of 2, also due to the
two-slope dependence of the CE01 conversion.

\subsection{Spitzer MIPS \70mu    imaging}

The \70mu     data used here are a combination of all of the GOODS-N
and GOODS-S available to us from the Spitzer programs ID-3325, -20147,
-81, and -30948.  The data were reduced following the techniques given
in Frayer et al.\ (2006).  These data represent 3~hours of integration over
the full GOODS-N field and the central
($10^{\prime}\times10^{\prime}$) region of GOODS-S.  The outer regions
of GOODS-S have an integration time of about 2~hours.  The typical
$3\sigma$ limit of these data is 1.8mJy at \70mu     in the deepest
regions.  Only 7 $BzK$ star forming $z\sim 2$ galaxies are individually
detected at \70mu    (6 of which are in GOODS-N, a likely effect of
small number statistics).  Therefore, we used stacking techniques to measure
average properties of galaxies at \70mu.  Bright low-redshift
sources were removed from the data to avoid biasing the stacking
results. We convert the \70mu    flux densities into total IR luminosities
again using the CE01 and DH02 template libraries. Being closer to the
peak of IR SEDs, the conversion of observed \70mu     flux densities into
$L_{\rm IR}$ is much less critical than for \24mu, even at $z\sim2$.

\subsection{VLA 1.4\,GHz radio imaging}

For this work, we use new ultradeep 20 cm (1.4\,GHz) imaging of the
GOODS-N obtained at the VLA in various array configurations for
maximizing the data quality and the dynamic range (Morrison et al., in
preparation).  Coadded to the previous radio data (mostly
from Richards et al.\ 2000), the new data currently reach an rms of
4.5\,$\mu$Jy per beam over the field. We then assume a radio spectral index
$\alpha=-0.8$ for deriving 1.4~GHz rest-frame radio luminosities from
the observed flux density. The local radio-IR correlation (Condon et al.\ 1992; Yun,
Reddy \& Condon 2001) is used to estimate the bolometric IR luminosity
from the radio luminosity:

\beq
L_{\rm IR}/L_\odot = 3.5\times10^{-12} L(1.4~GHz) \quad [W Hz^{-1}]
\label{eq:radioIR}
\eeq

Using radio luminosities to estimate $L_{\rm IR}$ for our sources implies
assuming that the radio-IR correlation still holds at high redshift.
The dispersion of the local radio-IR correlation can reach a factor of
2 for the most luminous sources (Yun et al.\ 2001), with luminosities
comparable to those of the objects detectable at $z\sim2$. The
introduction of a K-correction term, to derive 1.4~GHz rest-frame flux
densities from the observed ones, can introduce extra noise (or a
bias) if the radio slope is different from the assumed value.  This
effect is likely smaller than the intrinsic dispersion seen locally, as a
10\% error in the slope translate in a similar error in the intrinsic
1.4~GHz rest-frame flux density and therefore in the estimated
$L_{\rm IR}$.

In order to reach deeper radio flux densities, we stacked individually
undetected sources using the following technique. We measured the
average radio flux density level in a 1$''$ radius aperture and correlated
this measure to the total radio flux density using simulated radio
sources of various flux densities (defining an aperture correction, about a
factor of 1.4).  By measuring a large number of empty regions in the
radio image we estimated the rms fluctuations of this measure in order
to associate an error to each measure.  This allows us to recover
radio flux densities of ultra-faint sources with an rms of 4.7\,$\mu$Jy per
source, very close to the noise per beam.  We used a 1$''$ radius
aperture, very close to the radio beam size, after testing a variety
of apertures from 0.5$''$ to 2$''$, as it provides the best compromise
between final S/N and the necessity to measure over a relatively large
area in order to cope with possible coordinate misalignments. To
statistically account for the coordinate mismatches (of order of
0.1--0.2$''$). By applying this procedure to bright, individually
detected galaxies based on the K-band prior position, and comparing
the result to radio flux  densities derived by direct fit of the profiles, we
found that the radio flux densities required an average additional
correction of 10\%.  All these measurements and errors take into
account the primary beam correction for off-axis measurements. For
obtaining stacking measurements of undetected sources we obtained
weighted averages of the flux densities measured in this way.

\subsection{SCUBA 850\,$mu$m imaging}

The GOODS-N field has been observed with the Submillimeter Common User
Bolometer Array (SCUBA) at 850\,$\mu$m in a variety of programs, whose data 
have been merged together into a so-called super-map by Borys et al.\ (2003) 
and Pope et al.\ (2005).  The depth of the submillimeter (submm) map is highly 
variable over the field.  In D05b, we found that only one 
GOODS-N $BzK$ galaxy with $K<20$ is individually
detected at 850$\mu$m in the catalog 
described in Pope et al.\ (2006).  When considering $BzK$ galaxies down to 
$K < 20.5$, as in the present work, we detect five objects at 850$\mu$m.
We note again here that we are excluding all submm galaxies (SMGs)
that are individually 
detected in hard X-rays.

We obtained SCUBA 850\,$\mu$m stacking of individually undetected
sources by estimating flux densities and errors from the super-map at the
position of each program galaxy, and then proceeding as in D05b.

Using sub-mm flux densities at 850\,$\mu$m to infer bolometric luminosities
($L_{\rm IR}$)  relies on  assumptions for the long wavelength shape
of the SED, i.e., on dust temperatures. For this work, we continue to
use the SED shape templates of CE01 and DH02, derived from local
correlations.  The $f_{850\,\mu m}$ to $L_{\rm IR}$ conversion is to first
approximation independent of redshift for the $1.4<z<2.5$ range
explored here. However, the conversion is not perfectly linear.
According to the CE01 templates, a ULIRG ($L_{\rm IR}=10^{12}L_\odot$) would
have a flux density of about 1~mJy at these redshifts, raising to
about 4.6~mJy for a HyLIRG ($L_{\rm IR}=10^{13}L_\odot$). DH02 templates give
$\sim 20$\% higher flux densities at a given $L_{\rm IR}$, over the same regime
of luminosities.  The non linearity is due to the luminosity-temperature 
trend observed in the local Universe (less luminous
galaxies harbor colder dust).

\begin{figure*}
\centering
\includegraphics[width=18cm]{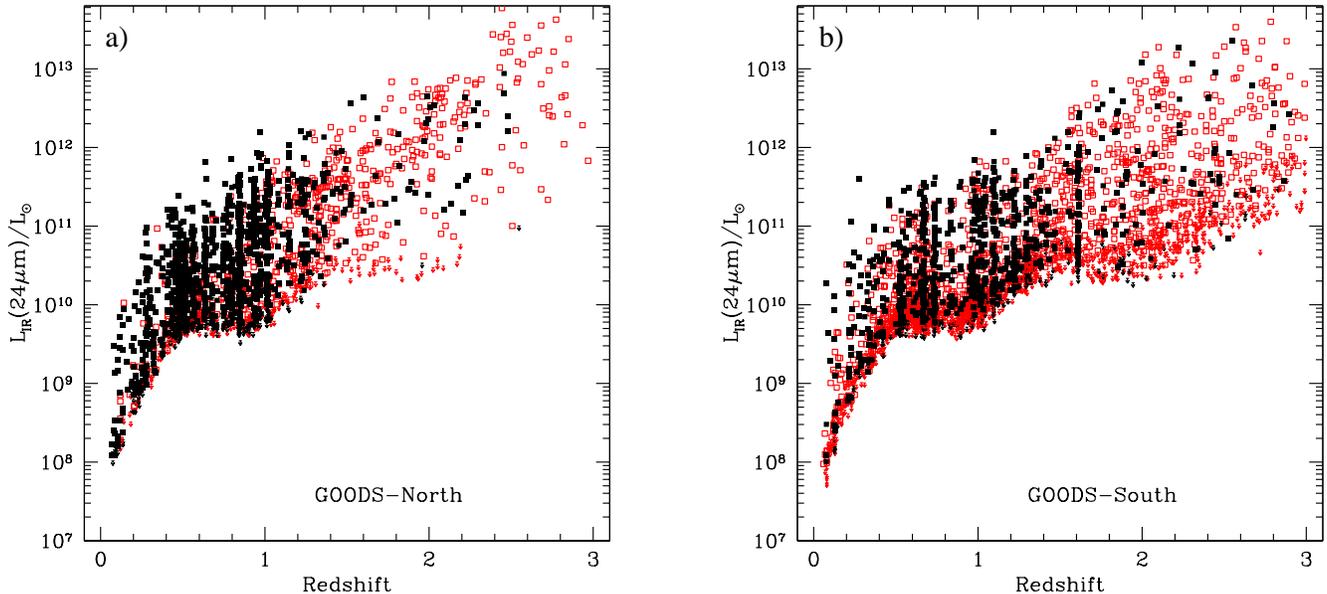}
\caption{
The bolometric infrared luminosity of galaxies $L_{\rm IR}$ estimated from the
\24mu    
flux densities using the templates of CE01 is plotted against redshift for $K<20.5$
 sources in GOODS-N (left panel) and GOODS-S (right panel). Symbols are as
in Fig.~\ref{fig:24umLIR_N}.
}
\label{fig:24um_N}
\end{figure*}

\subsection{X-ray imaging}

We use the catalog of X-ray detected sources of Alexander et al.\
(2003). All sources detected in the
hard X-ray band were excluded from the SFR analysis in this paper as
likely containing AGNs. SFRs can be derived from the soft band 0.5--2~keV
fluxes (mapping closely to the 2--8~keV rest-frame band) at $z=2$
(see e.g., Ranalli et al.\ 2004,
Persic et al.\ 2004, Hornschemeier et al.\ 2005), provided that no
substantial AGN contamination is present at those energies.
We have performed X-ray stacking of
$BzK$ samples over the two fields. Results are presented in \hbox{Paper~II},
as an assessment of the presence of AGNs is required for interpreting
the X-ray stacking results.

\subsection{UV based SFR estimates}
\label{UV_S}

Ongoing SFRs in $z\sim2$ galaxies are estimated using the UV
luminosity, following the prescriptions in D04b. The
1500\AA\ rest-frame luminosity is derived from the observed $B$-band
flux density, and a K-correction term is derived based on the redshift of each
source (photometric or spectroscopic) and its UV spectral slope. Then
L(1500\AA) is converted into a SFR using Eq.~5 of D04b, i.e.:

\beq
{\rm SFR} (M_\odot{\rm yr}^{-1}) = L_{1500 \AA} [{\rm erg\ s^{-1}
Hz^{-1}}] / (8.85\times10^{27}).
\label{eq:madau}
\eeq

\noindent
 The amount of dust reddening is estimated adopting a a Calzetti et
 al.\ (2000) reddening law (see also Meurer, Heckman \& Calzetti  1998), and using the relation from D04b:

\beq
E(B-V)=0.25(B-z+0.1)_{AB}.
\label{eq:redde}
\eeq

\noindent
An attenuation of $A_{1500}=10\times E(B-V)$ mag is then adopted to obtain
dust-corrected SFRs from Eq.~8.
With this procedure we obtain reddening corrected SFRs consistent to what 
derived with SED fitting (D04b) for the bulk of
$K$-selected $z\sim2$ galaxies.

Clearly, this technique is affected by several limitations.  In
applying a reddening correction (which median is a factor of $\sim 40$
for $K<20$, $z=2$ galaxies), the technique assumes that the UV
spectral slope is entirely due to reddening, rather then to the
presence of evolved stellar populations. Therefore we expect it should
only work for actively star forming galaxies.  Even more importantly,
it is well known that strongest starbursts are opaque to UV radiation
and its total SFR activity cannot be reliably estimated solely from
the UV, even after reddening corrections (e.g., Goldader et al.\
2002; Buat et al.\ 2005). 
Things can be different though at higher redshifts, as shown
later in the paper (see also Flores et al. 2004; D05b; Reddy et al. 2005).

\subsection{IMF related effects}

It is worth mentioning to which extent the analysis presented in this
paper depends on the adopted IMF, i.e., a Salpeter (1955) IMF from 0.1
to 100 $M_\odot$. All of the luminosities that we use as SFR
indicators are sensitive to the population of high mass stars
($\simgt5M_\odot$), but the bulk of the mass in stars that is actually
formed is dominated by lower mass stars. Therefore, changing the slope
of the low-mass end of the IMF, e.g., adopting the IMF of Kroupa (2001)
or Chabrier (2003) that are in better agreement with current
observations, would just imply lower overall SFRs, i.e., different
normalizations in \hbox{Eqs.~1~and~7} and different stellar masses (by
nearly the same factor). We prefer to quote numbers for Salpeter IMF
for consistency with much of the literature in the field.

However, changing the intermediate or high-mass slope could have a
significant effect on the relative conversions between different SF
indicators, since stars in different mass ranges can be responsible
for varying contributions to the different SF indicators. For
example, the radio emission is thought to be powered by higher mass
stars than the IR or the UV emission (e.g., Condon et al.\ 1992), and a
steeper IMF slope at the high masses would systematically increase the
IR/UV over radio luminosities. Timescale issues, e.g., observing a
starburst very close to the beginning (Roussel et al.\ 2003) or a long
time after the SFR peak, can also alter the luminosity ratios.
Conceivably, IMF and star formation timescale are partly degenerate
for some of the effects that we discuss later in the
paper. We neglect timescale effects in the following of the paper.

\subsection{Limiting depths and effect of photometric redshifts}

Fig.~\ref{fig:pl_limi} shows the limiting sensitivity to bolometric
IR luminosity from the various tracers using the recipes described in
the previous sections (deep radio and SCUBA photometry are available
only in GOODS-N).  It is clear that the MIPS \24mu    flux density is the most
sensitive IR tracer by at least an order of magnitude, allowing in
GOODS to detect galaxies with $L_{\rm IR}\sim10^{11}L_\odot$ for most of
the $1.4<z<2.5$ redshift range considered here. This is illustrated in
Fig.~\ref{fig:Det_Rate}, showing that the fraction of $z\sim2$
galaxies in our sample detected at \24mu    remains very high, over
60\% even at the $K<22$ limit of the GOODS-S field.  Even lower SFRs
can in principle be derived from the UV over the whole redshift range,
provided there is no or just small dust obscuration. An unreddened galaxy
at $z=2$ with $B=27$ AB (3$\sigma$ limit) would
have SFR\,$=0.6$~M$_\odot$~yr$^{-1}$.
A more practical limit is given by the faintest galaxy
detectable in the $K$-band ($K=22$) having no dust reddening, 
which would have SFR\,$\sim10\ M_\odot$~yr$^{-1}$ for
$z=2$, similar to the depth reached at \24mu.    In the case of
galaxies with non zero dust reddening we can measure the SFRs only above
higher limits. For example, a galaxy with $B=27$ and $E(B-V)=0.4$ (the
median value at the bright $K<20$ limits) would have SFR\,$\sim25$
M$_\odot$~yr$^{-1}$ at $z=2$ (Fig.~\ref{fig:pl_limi}).  It is
therefore quite interesting to compare UV and \24mu    based SFR and
$L_{\rm IR}$ estimates, given that these are the most powerful tracers,
and available for nearly all of the $z\sim2$ galaxies in our GOODS sample.

The dependence of $L_{\rm IR}$ with redshift is generally steep, and
varying as a function of redshift itself, except for SCUBA 850\,$\mu$m
photometry. In turn, photometric redshift errors propagate into
$L_{\rm IR}$ and
SFR errors differently when different tracers are used.  We
note, however, that this is mitigated when considering the ratios of
$L_{\rm IR}$ derived from two independent tracers, as is generally done in
this paper.  Even in this case, however, one can commit significant
errors in cases of catastrophic redshift failures, e.g., if a galaxy
at $z < 2$ is mistaken to be at $z \sim 3$.  In this case, the
$L_{\rm IR}$ derived from the \24mu    or \70mu    data could be strongly
overestimated relative to that derived from, e.g., 20~cm data or from
the UV, due to the stronger K-corrections for the emission in the MIPS
bandpasses.  We take these effects into account when comparing the
SFRs from different tracers. In general, the high quality of the
adopted photometric redshift (see Fig.~\ref{fig:zphot_N}) should
limit this problem to a small minority of cases. In addition, we use
the extensive spectroscopic redshift sample to consolidate any
conclusion first derived using the full combination of spectroscopic
and photometric redshifts.

\begin{figure}
\centering
\includegraphics[width=8.8cm]{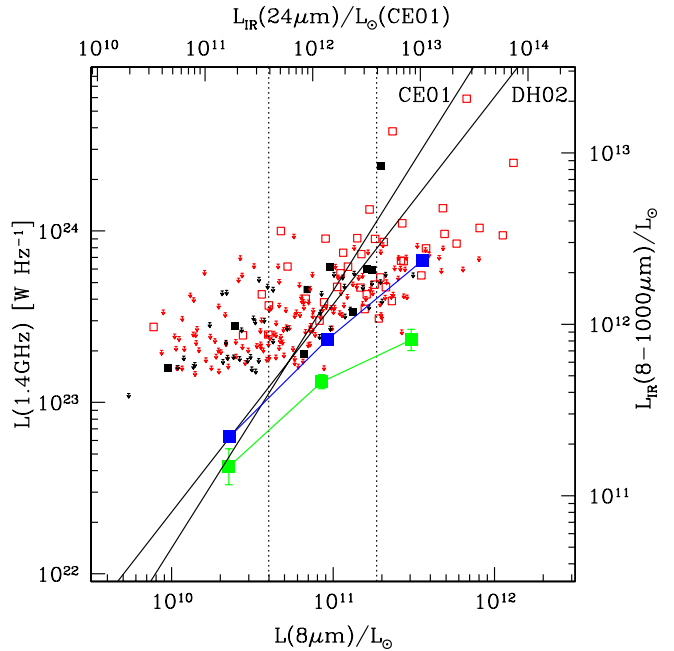}
\caption{Comparison, for $z\sim2$ galaxies in the GOODS-N field, of the total IR luminosity ($L_{\rm IR}$) as inferred from radio (using the local 
radio-$L_{\rm IR}$ correlation; Condon 1992; Yun et al.\ 2001) 
to the 8\,$\mu$m rest-frame luminosity (derived from the \24mu    observed
flux density). Black and  red symbols are as in Fig.~\ref{fig:24umLIR_N}.
The green points show the results of stacking
radio undetected sources in 3 bins of 8\,$\mu$m luminosity.
The blue points show the average
trend versus 8\,$\mu$m luminosity, including both radio detected and undetected sources.
The solid lines show the expected correlation based on the CE01 and DH02 templates.
Hard X-ray detected AGNs are excluded.
}
\label{fig:xlum_N}
\end{figure}

\begin{figure*}
\centering
\includegraphics[width=18cm]{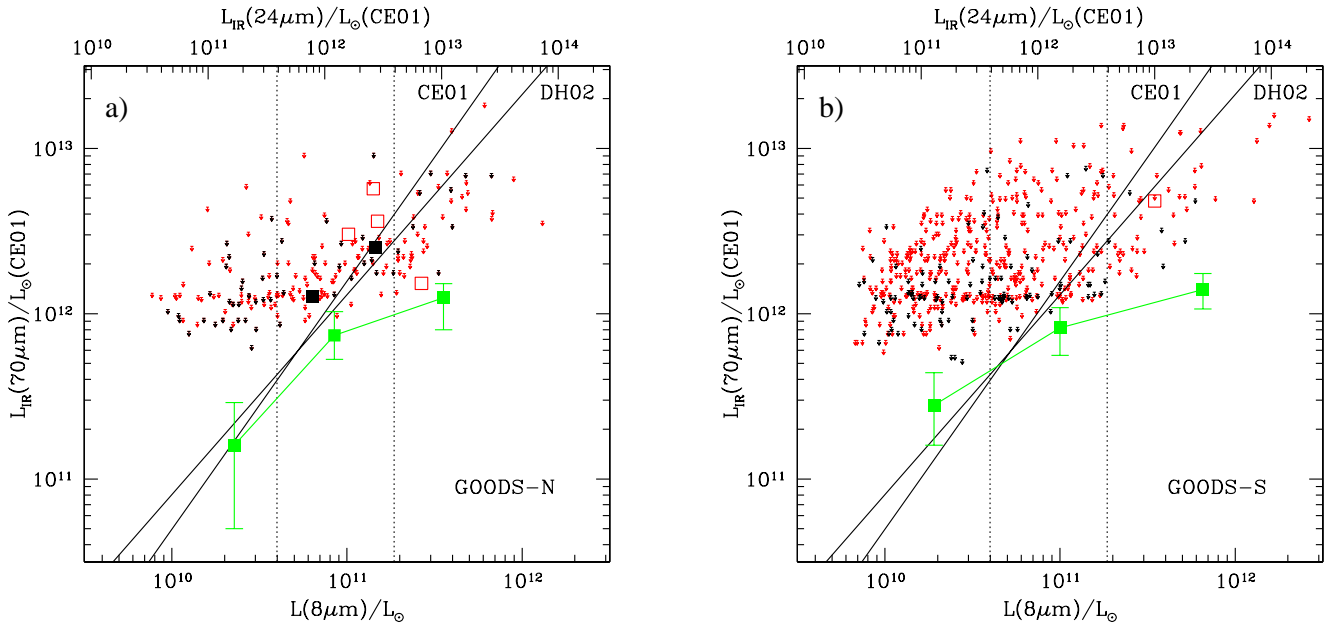}
\caption{
Comparison, for $z\sim2$ galaxies, of the total IR luminosity ($L_{\rm IR}$), or limits, as inferred from \70mu    
Spitzer photometry to the 8\,$\mu$m rest-frame luminosity derived from MIPS \24mu. The left panel is for GOODS-N, while the right panel is for GOODS-S.
The solid lines show the expected correlation based on the CE01 and DH02 templates.
Black and red squares are for individual detections, symbols are as in Fig.~\ref{fig:24umLIR_N}.
The green point show the results of stacking
\70mu    undetected sources in 3 bins of 8\,$\mu$m luminosity.
Hard X-ray detected AGNs are excluded.
}
\label{fig:xlum70_N}
\end{figure*}

\begin{figure}
\centering
\includegraphics[width=8.8cm]{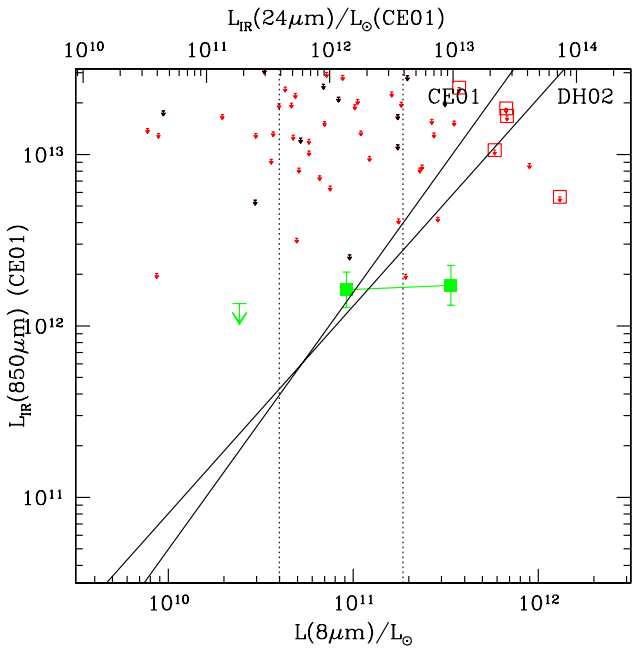}
\caption{
The comparison of SCUBA 850\,$\mu$m inferred $L_{\rm IR}$ for $z\sim2$ objects to the \24mu    based
luminosities. For the 5 SMGs (red squares)
we have photometric redshifts $2.2<z<2.8$.
The green points show stacking of 850\,$\mu$m undetected sources.
Hard X-ray detected AGNs are excluded.
}
\label{fig:xlum850_N}
\end{figure}

\section{IR Bolometric luminosities from mid-IR, sub-mm and
radio}\label{sec:mips}

\subsection{Bolometric luminosities from the \24mu    flux densities}

Fig.~\ref{fig:24umLIR_N} shows the flux density versus redshift relation of
MIPS \24mu    detected sources in the two GOODS fields, to the
corresponding $K$-band limit in each field. Fig.~\ref{fig:24um_N}
shows the bolometric IR luminosity inferred from the \24mu flux densities
using the CE01 templates. From both sets of figures it is apparent
that moving to higher redshifts sources with brighter and brighter
$L_{\rm IR}$ are found, implying (after accounting for volume effects)
a rapid brightening of the IR luminosity
function (as already emphasized in the literature; e.g.,
Le Floc'h et al.\ 2005; Caputi et al.\ 2007).
However, it is somewhat surprising that the rise extends all the way
to redshift $z\sim3$, and sources with $L_{\rm IR}\sim10^{13}L_\odot$ and
even $L_{\rm IR}\sim10^{14}L_\odot$ appear in the GOODS fields (see Peres
Gonzales et al.\ 2004; Papovich et al.\ 2006; Caputi et al.\ 2006 for
similar results).  If due to star-formation, these extreme
luminosities would imply obscured SFRs up to 10000 $M_\odot$
yr$^{-1}$. Such extremely high SFRs as inferred from \24mu flux densities
are somewhat suspicious, suggesting that the mid-IR light may not be
entirely powered by star formation, or may be reflect
different intrinsic mid-IR properties of these $z=2$ star forming galaxies, 
with respect to local sources.

\subsection{MIPS \24mu versus radio luminosities}

Bolometric luminosities derived from the \24mu flux densities are
compared to those derived from  radio in Fig.~\ref{fig:xlum_N}, for GOODS-N
$z\sim2$ galaxies. For radio detected sources (about 21\% of the total
sample to $K=20.5$) the average galaxy has
$L_{\rm IR}\sim10^{12}L_\odot$ and there is overall agreement between
\24mu and radio derived luminosities, for both CE01 and DH02
templates. However, a large scatter is present and there are galaxies
with \24mu    to radio inferred $L_{\rm IR}$ ratios much different from
unity in both directions.  We will come back to these sources later in
this section.  For 94\% (46/49) of the galaxies with $L(8\mu
m)>2\times10^{11}L_\odot$, the radio detection (or limit) implies a
higher  $L_{\rm IR}$ estimated from \24mu    with respect to radio, using
either CE01 or DH02 templates.

We used radio stacking in order to investigate trends for
radio-undetected sources. Green symbols in Fig.~\ref{fig:xlum_N}
show the result for radio-undetected sources, stacked as a function of
the 8\,$\mu$m rest-frame luminosity, in turn derived from the \24mu   
flux density. Stacked flux densities are in the range of 3--7\,$\mu$Jy, comparable to
or larger than the single beam noise, and the signal to noise of the
detections is larger than 3 in all cases.  The stacking emphasizes the
existence of a substantial population of sources with $L(8\mu
m)>2\times10^{11}L_\odot$ where the $L_{\rm IR}$ estimated
from \24mu    exceeds on
average by an order of magnitude the same quantity derived from
radio. When considering together radio detections and non-detections, we
see a good agreement at low $L(8\mu m)$ luminosities and an overall
increasing discrepancy of higher $L(8\mu m)/L_{\rm IR}$ with respect to
 both CE01 and DH02 templates.  For $L(8\mu m)>2\times10^{11}L_\odot$
[$L_{\rm IR}(24\,\mu m)>4\times10^{12}L_\odot$], the average overestimate is
about a factor of 3.  It can be noticed from Fig.~\ref{fig:xlum_N}
that radio limits move to higher $L_{\rm IR}$ in the right part of the
plot. This corresponds to the fact that most of these putative HyLIRGs
(from \24mu) tend to lie at higher redshift than average.

There are eight galaxies, spread over a range of mid-IR and radio
luminosities, that on the contrary show a radio-derived bolometric
luminosity that exceeds by a factor $\simgt 3$ (and up to $\sim30$)
that inferred from $L(8\mu m)$.  This could be due to additional radio
emission from an AGN (despite non detection at X-ray wavelengths), as
suggested also by Donley et al.\ (2007), mostly in lower redshift
sources. Some sources might have \24mu    {\em deficit}, as expected
in the case of very low metallicity (Engelbracht et al.\ 2005; Madden et
al.\ 2006), or in the case of a strong 9.7 $\mu$m silicate absorption
entering the \24mu    bandpass. The latter case can happen for
$z\sim1.5$, and we do find that 6 out of the 8 galaxies are indeed at
$1.2<z<1.8$.

\subsection{MIPS \24mu    versus MIPS \70mu}

We have cross-checked this result using additional $L_{\rm IR}$
indicators. Fig.~\ref{fig:xlum70_N} panels show the comparison with
luminosities inferred from the MIPS \70mu    flux densities from the survey of
Frayer et al.\ (2006; and in preparation).  In the area with available
deep \70mu    imaging from Spitzer, about 200 arcmin$^2$ from the two
fields, there are only a handful of individually detected galaxies at
$z\simgt1.4$.  The \70mu     imaging has reached much shallower
$L_{\rm IR}$ limits than the \24mu    data; however the non-detections
are meaningful for the sources with higher $L_{\rm IR}$ as
inferred from the \24mu    data and imply
that $L_{\rm IR}$ from \24mu    is overestimated in those
sources.  Similarly to radio, we have performed stacking of galaxies
at \70mu     in both GOODS fields, binning the individually undetected
galaxies as a function of their 8\,$\mu$m rest-frame luminosities (see
green boxes in Fig.~\ref{fig:xlum70_N}). Again, the most notable
feature is that sources with the brightest 8\,$\mu$m rest-frame
luminosities are not correspondingly brighter at \70mu, as would
be expected on the basis of local templates. The average \70mu     to
 \24mu    flux density ratio is $\sim 6-7$ for sources with $L(8\mu
m)\sim10^{11}L_\odot$, and of$\sim 3$ for those with $L(8\mu
m)>2\times10^{11}L_\odot$.  Templates from CE01 and DH02 predict instead
that the \70mu     to  \24mu    flux density ratio at $z\sim2$ should steeply
increase with intrinsic $L_{\rm IR}$. A similar trend is found by Papovich
by al. (2007) in $z=2$ galaxies.

Notice that contrarily to the radio case, for a given $L_{\rm IR}$ the
\70mu    to \24mu    flux density ratio is fairly independent on the redshift, for
a given template, in the range considered including $2<z<3$, due to
more similar K-corrections.  Therefore these conclusions are
relatively robust against uncertainties coming from photometric
redshifts.

\begin{figure}
\centering
\includegraphics[width=8.8cm]{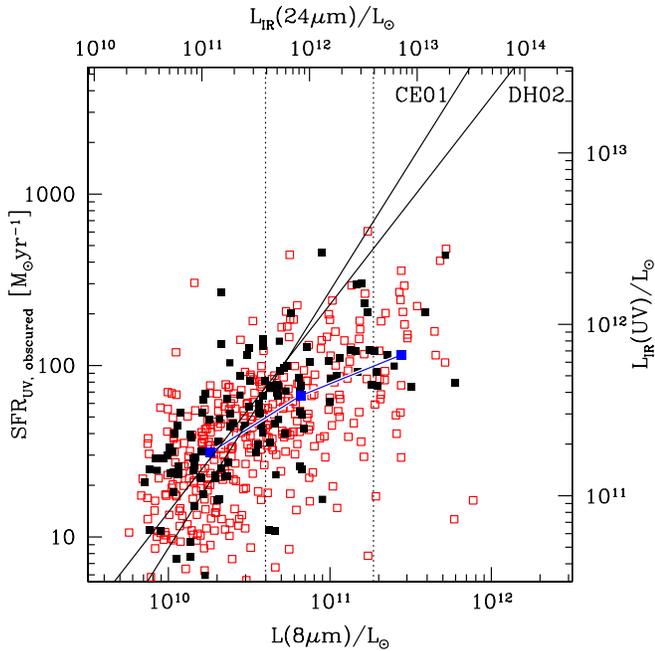}
\caption{The \24mu    inferred luminosities are compared to the UV light
from star-formation
reprocessed into the IR (SFR$_{UV, obscured}$). 
The latter is defined as the difference between
the dust-corrected and the
uncorrected SFR estimated from the 1500\AA\ rest-frame luminosities. The
blue points show the behavior
of the median galaxy, for the usual 3 bins of $8\mu$m rest-frame luminosity.
Hard X-ray detected AGNs are excluded.
}
\label{fig:xxx_uv_S}
\end{figure}

\subsection{MIPS \24mu versus SCUBA 850\,$\mu$m}

Finally, we have used the SCUBA super-map from Borys et al.\ (2003) in
order to investigate the comparison of 850\,$\mu$m versus  \24mu   
inferred $L_{\rm IR}$ values for $z\sim2$ galaxies. Only 5 galaxies in the
sample are individually detected, all at the highest luminosities. For
4 of them the bolometric luminosities inferred from the  \24mu   and
850\,$\mu$m flux densities are in good agreement with expectations based on local
templates,
while in a single case the  \24mu    flux density lead to overpredicting
$L_{\rm IR}$ by over an
order of magnitude compared to SCUBA.  It is interesting to notice
that this result is different from the findings by Pope et al.\ (2006)
for 850\,$\mu$m selected galaxies, where for typical galaxies the
850\,$\mu$m flux density was in excess of what could have been predicted from the
 \24mu    flux density. Our 5 sub-mm detected $BzK$ galaxies represents actually a
small part of the Pope et al.\ (2006) sample. The most likely
explanation of the differences is that the sub-mm selection favors galaxies
with relatively cold dust, as emphasized also in Pope et al.\ (2006).

We stacked 850\,$\mu$m undetected sources as a function of their $8\;\mu$m
luminosities. We detect the intermediate and highest luminosity bins
only, both with S/N=4 and $S_{850\,\mu m}\sim1.3$mJy. This is consistent
with the result presented in D05b and Knudsen et al.\
(2005) for SCUBA stacking of distant $z\sim2$ sources. This also
confirms the result that for the most luminous 8\,$\mu$m rest-frame
galaxies, the 8\,$\mu$m luminosity is in excess of what the local
templates would
predict given the $L_{\rm IR}$ of the galaxies estimated from other
means. As to first order the 850\,$\mu$m flux density is independent of
redshifts for IR luminous galaxies within the probed redshift range, finding
the same 850\,$\mu$m for the two bins shown in Fig.~9 while their
 average $8\mu$m luminosity differers by
a factor of $\sim 4$ testifies again for a large scatter in $8\;\mu$m to
bolometric $L_{\rm IR}$ luminosity ratio for high-$z$ LIRGs and ULIRGs in our
K-selected
sample.

\section{SFR from UV light: 'transparent' ULIRGs at $z\sim 2$}
\label{sec:uv}

D05b found that the UV estimated SFRs, averaged over
the full sample of $BzK$ star forming galaxies with $K<20$, was well
within a factor of 2 of the average of the estimates derived from the
radio, mid-IR, X-rays and sub-mm data.  This suggested that --at least
on average-- UV is a a good tracer of star-formation for massive $z=2$
galaxies, as advocated e.g., also by Reddy et al.\  (2005).  We investigate
here to which extent this remains valid when considering individual
galaxies.

\subsection{UV versus MIPS \24mu}

Fig.~\ref{fig:xxx_uv_S} shows the UV estimate of the
{\it obscured} SFR as a function of
the 8\,$\mu$m rest-frame luminosities, for the GOODS-S
field only where we reach much fainter star-forming galaxies.
This quantity is defined as:

\beq
{\rm SFR}_{\rm UV, obscured} = {\rm SFR}_{\rm UV, corr} - {\rm SFR}_{\rm UV, uncorr},
\eeq where the correction is for dust extinction, as
described in Section~3.6. The term SFR$_{\rm UV, obscured}$ is equivalent to 
SFR$_{\rm IR}$ in \hbox{Eqs.~1 and~2}.
Moreover, entering with this quantity 
into Eq.~1 and solving for $L_{\rm IR}$ one gets the expected
infrared luminosity as due to the partial obscuration of the star
formation and the reprocessing of its UV photons, i.e., $L_{\rm IR}(UV)$,
also plotted in Fig.~\ref{fig:xxx_uv_S}. For this comparison we limit
to sources detected at \24mu     and to galaxies where the
$(B-z)_{AB}$ color is well estimated, with total error below 0.4~mag
(corresponding to a maximum acceptable error of a factor of 2.5 in the
reddening correction to the SFR).  At low 8\,$\mu$m rest-frame
luminosities, there is a fairly good agreement between IR light (and
SFR) estimated from UV and from the \24mu flux density, which is within a
factor of 2 for most galaxies. The two estimates start to diverge for
$L(8\mu m)\sim 4\times10^{10}L_\odot$, and by $L(8\mu
m)>2\times10^{11}L_\odot$ the UV underestimates the IR light by more
than a factor of $\sim 6$ compared to the mid-IR. However, given the
discussion in the previous section, it is not clear how often this is
due to an underestimate of the obscuration of the UV light, or to an
excess emission at \24mu.     

\begin{figure}
\centering
\includegraphics[width=8.8cm]{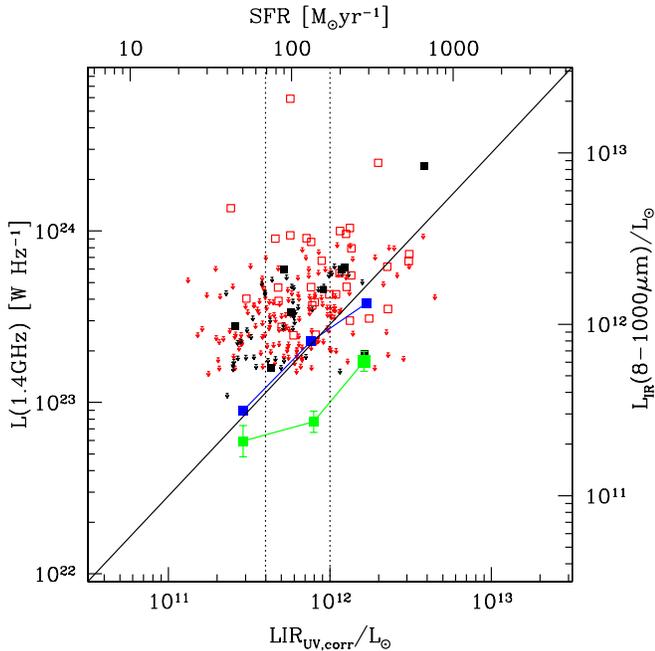}
\caption{The comparison of radio to UV estimated total IR luminosities.
Green points show the stacking
of radio undetected sources in the 3 bins defined by the dotted vertical
lines. Blue points show the
average of all galaxies (including detections) in the same 3 bins.
Hard X-ray detected AGNs are excluded.
}
\label{fig:rad_lum_N}
\end{figure}

\subsection{UV versus radio}

In order to further check the reliability UV-derived SFRs we compared
UV-based estimates to radio observations.  For radio detected sources
(20\% of the sample) we find that the median ratio between the IR
luminosity estimated from the radio to that from the UV is about a
factor of 2, as shown in Fig.~\ref{fig:rad_lum_N}.  This is likely
the effect of the limiting flux density in the radio, such that only the
sources with brightest continuum are detected (i.e., the Malmquist
bias).  When stacking radio-undetected sources as a function of UV
inferred $L_{\rm IR}$, and considering together radio-detected and
undetected sources, Fig.~\ref{fig:rad_lum_N} shows that the ratio of
UV to radio $L_{\rm IR}$'s appear to be very close to unity at all UV
luminosities, within the errors.  This implies that the UV-based SFR,
corrected
for dust reddening, is on average a good tracer of SFR even for the
case of ULIRGs with $L_{\rm IR}\simgt10^{12}L_\odot$.

On the other hand, there are also evidences of extreme obscuration at UV
wavelengths. For 2-3 sources in the sample the UV underestimates
$L_{\rm IR}$ by a factor up to 10--30
(some of these objects
might be radio loud AGN, despite their non detection in the X-ray).
For the most luminous sources with $L_{\rm IR}>2\times10^{12}$ based on
radio (a limit above which we should have a complete sample over
GOODS-N) the median galaxy has UV underestimating $L_{\rm IR}$ by a factor
of 2.7. For the typical (or median) ULIRG in our sample, however, the
UV appears to provide a fairly good estimate of $L_{\rm IR}$, with no
systematic trend/offset with respect to the radio estimates. This is
quite different from local ULIRGs, where the UV is known to
underestimate $L_{\rm IR}$ by factors $\simgt10$ (Trentham et al.\ 1999; Goldader
et al.\ 2002) even after correcting for dust extinction.

Similarly, for $z\sim2$ SMGs the UV underestimates
SFRs by an average factor of 120 (Chapman et al.\ 2005).  For the 5
galaxies in GOODS-N to $K=20.5$ that are detected by SCUBA at
850\,$\mu$m, the 3 objects where the UV slope can be estimated with
relatively good S/N the ratio of radio- to UV-derived  SFR is in the range
between 3 and 35. For the other two objects, with a poorer quality UV-derived
 SFR, this ratio is also $\sim 30$. As we are considering here a
subsample of SMGs with bright near-IR magnitudes, it is likely
that we are limiting to the least obscured sources in their class,
hence these results appear to be consistent with Chapman et al.\ (2005).

\begin{figure}
\centering
\includegraphics[width=8.8cm]{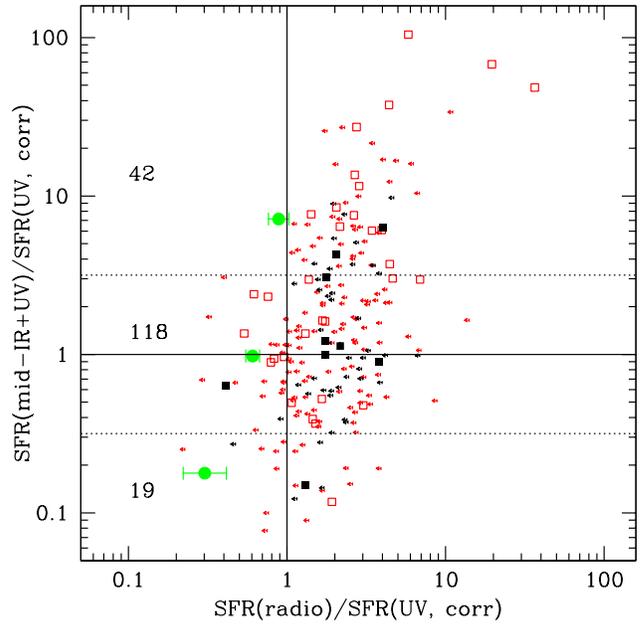}
\caption{The ratio of mid-IR and radio inferred SFRs to UV inferred ones.
Green points are the stacking
of radio undetected sources in the 3 bins defined by dotted lines.
Number of sources stacked in each bin is reported.
Hard X-ray detected AGNs are excluded.
}
\label{fig:who_right}
\end{figure}

All in all, there seems to be evidence that the typical $z\sim2$ ULIRG
is fairly 'transparent' to UV light, the word meaning that we can 
estimate its SFR given its
rest-frame UV 1500\AA\ luminosity and UV slope. However, at much brighter
bolometric luminosities, well in excess of $10^{12}L_\odot$, the UV
starts to saturate, as inferred from the brightest radio or sub-mm
galaxies. There appears to be a shift of UV saturation level from
about $L_{\rm IR}=10^{11}L_\odot$ at $z=0$ to $L_{\rm IR}$ greater than a few
$10^{12}L_\odot$ at $z=2$. This is in line with the overall rise of
the SFR density of the Universe during these epochs.

Thus, it appears that the UV is on average performing well for the
$L_{\rm IR}\simlt10^{12}L_\odot$ $z=2$ galaxies. However, for the
discrepant cases with $L_{\rm IR}(24\,\mu{\rm m}) \gg L_{\rm IR}(UV)$ (see again
Fig.~\ref{fig:xxx_uv_S}) it remains unclear whether the discrepancy
is due to excess obscuration or to an intrinsic \24mu    excess. To
disentangle this ambiguity we have again used the radio as a
diagnostic tool. In Fig.~\ref{fig:who_right} we show the total SFR
[from Eq.~2 using $L_{\rm IR}(24\,\mu{\rm m})$ as input into
Eq.~1] of the radio-derived SFRs, both normalized to the
reddening corrected SFR from the UV.  For the sources whose SFR
derived from the \24mu flux density exceeds by more of a factor of 3 that
derived from the UV, we find that only $\sim10$--15\% of them have a
similar or larger excess also in the radio (and thus could be
represent ULIRGs with anomalously high obscuration). For the majority
of \24mu     excess objects, however, we find again that radio SFRs
agree with UV SFRs, suggesting again that it is \24mu     flux density to be
genuinely in excess also with respect to the UV luminosity.

\begin{figure}
\centering
\includegraphics[width=8.8cm]{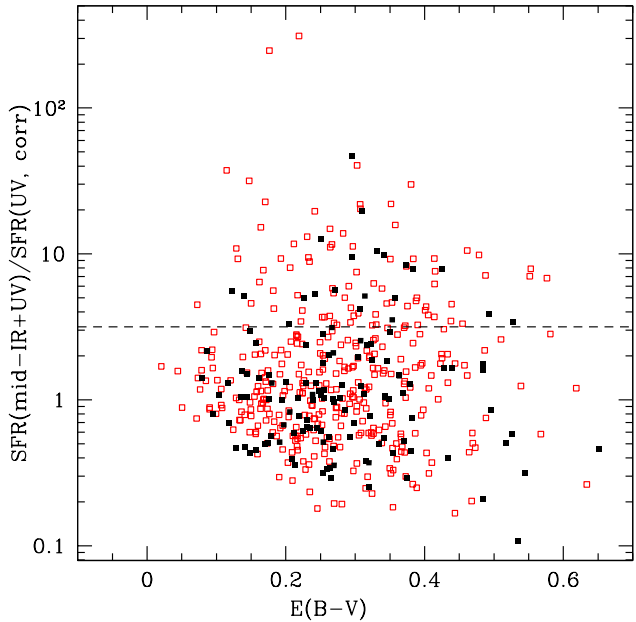}
\caption{
The ratio of SFR inferred from \24mu (plus the UV uncorrected one) to the SFR
derived from UV only and corrected
for dust extinction using a Calzetti law, is plotted as a function of the
reddening E(B-V) inferred
from the UV slope (through the $(B-z)_{AB}$ color, as defined in D04b). 
We classify galaxies above the dashed horizontal line as {\em
mid-IR excess} galaxies. The nature of such sources is investigated in
detail in \hbox{Paper~II}.
}
\label{fig:EBV_S}
\end{figure}

\subsection{Reliability of reddening correction and quiescent galaxies}

The discrepancies that have emerged from the comparisons in the
previous sections lead to the question of how well we can correct UV
luminosities for dust reddening, given the adopted reddening law,
and if the mid-IR excess is somewhat related to dust reddening and UV slopes.
Fig.~\ref{fig:EBV_S} shows the ratio of ``total'' SFR as from
Eq.~2 estimated from the mid-IR using the \24mu flux densities
over the UV-corrected SFR, as a function of the estimated
reddening from Eq.~8. There appears to be no appreciable trend
of a larger such ratio as a function of reddening. This implies that
if excess obscuration is present over what corrected using the
Calzetti et al.\ (2000) law, then the UV slope has to be very poorly
correlated with such an excess obscuration. A {\it grey} obscuration
would work, but all known dust absorbers are not grey. On the other
hand, for the average galaxy, correcting the observed UV luminosities
based on the UV slope and using the Calzetti et al.\ reddening law,
allows us to obtain fairly robust SFR estimates, as indeed illustrated
by Fig.~\ref{fig:EBV_S}.

There are a few galaxies where the reddening corrected SFR from the UV
exceeds  that derived from the mid-IR by more than a factor of 3.  As opposed
to \24mu     
excess objects, Fig.~\ref{fig:who_right} shows that in most of these
cases it is the UV that is overestimating SFRs.  These latter ones are
mostly confined to the sources with reddest colors, and are
likely galaxies in a post-starburst phase that are on their way to turn
passive/quiescent. In these cases, the redness of the UV continuum is
due to a dearth of young/massive stars rather than to dust reddening, and
the SFR inferred from the UV, corrected for dust extinction
following the recipes of Section~3.6, is
obviously overestimated. These sources are intermediate cases between
passive $BzK$ galaxies and star-forming ones, and similar to the two
blue systems discussed in Daddi et al. (2005a). The fraction of these
systems is fairly small, implying a rapid transition  from the
star-forming stage to the passive one.  This is related to the
possible presence inside $sBzK$ samples of sources with passive or
quiescent stellar populations, as emphasized by Grazian, Nonino \&
Gallozzi (2007) and Quadri et al.\ (2007).  The 25\% discrepancy between
average radio and UV estimates of $L_{\rm IR}$ for the most luminous UV
galaxies (see the blue points in Fig.~\ref{fig:rad_lum_N}) is
completely solved by excluding these nearly quiescent sources (about
20\% of the radio undetected galaxies in that bin). Clearly, MIPS
\24mu     photometry is required to identify these likely {\em post starburst}
galaxies, where the UV corrected for dust extinction provides  an
incorrect estimate of the ongoing SFR activity.

\subsection{The nature of mid-IR excess galaxies}

The results of the previous sections have clearly singled out the
existence of a population of $z\sim2$ galaxies that we call mid-IR
excess galaxies, for which the MIPS \24mu     flux densities would imply IR
bolometric luminosities and SFR in excees to those implied by all
other available tracers of star formation activity. We emphasize 
that the existence of these mid-IR excess galaxies is independent on
the classes of templates used or  on the particular recipe to convert
an observed \24mu flux density at $z\sim2$ into a measure of SFR/$L_{\rm IR}$.
The result shown in \hbox{Figs.~7--10} can be read as evidence that
at a fixed bolometric luminosity around
$L_{\rm IR}\approx10^{12}L_\odot$ there is a very large scatter of
$L(8\,\mu{\rm m})$, and one can objectively define mid-IR excess galaxies as
the galaxies with the largest observed $L(8\,\mu{\rm m})$ 
for a givel $L_{\rm IR}$.

Understanding and investigating the nature of these sources and the
origin of the mid-IR excess has potentially far reaching
implications. These aspects are directly developed in full in 
\hbox{Paper~II}. In such paper we provide
evidence indicating that mid-IR excess galaxies contain heavily
obscured, often Compton Thick, AGNs. We then argue that the AGN
energetic input is the most likely cause of the mid-IR excess in these
galaxies.

\begin{figure*}    
\centering 
\includegraphics[width=18cm]{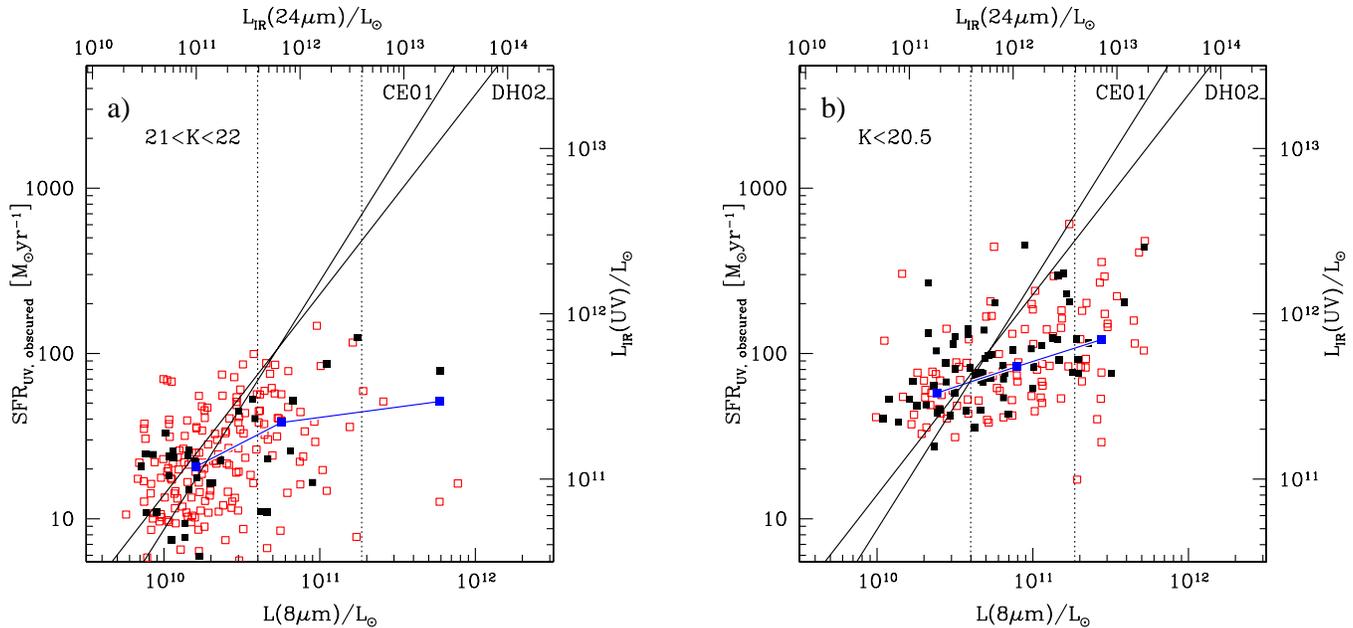}
\caption{The same of Fig.~10, but distinguishing $K$-band faint (left
panel) and bright (right panel)
galaxies. A clear shift in the mean trend between mid-IR luminosity and 
SFR is detected.
}
\label{fig:corre}
\end{figure*}

\section{Star formation rates in $z=2$ galaxies: a recipe}
\label{sec:recipe}

We summarize the work described in the previous sections by suggesting
a recipe for obtaining the best estimates of SFRs for samples of
$z\sim 2$ galaxies, by exploiting a variety of multiwavelength SFR
indicators as available in GOODS (and many of the current surveys,
although to less deep levels, such as e.g., in COSMOS, Scoville et
al.\ 2007). This procedure assumes that galaxies hosting relatively
unobscured and luminous AGN have been eliminated from the sample,
e.g,. by means of deep X-ray imaging.

\begin{itemize}
\item[1)] Compare UV-derived SFRs (SFR$(UV, corr)$), obtained as
described in Section~\ref{UV_S}, to the ones obtained from the 24 $\mu$m
flux density, using the CE01 or DH02 templates. For the sources where
SFR$(mid-IR+UV)$ and SFR$(UV, corr)$ agree within a factor of $\sim3$
(Fig.~\ref{fig:who_right}), use the 24 $\mu$m based SFR plus the
UV-based SFR uncorrected for reddening (i.e., SFR$(mid-IR+UV)$) as the
best estimate for the galaxy SFR. These sources are called {\em mid-IR
normal} star forming galaxies.

\item[2)] For the sources where SFR$(mid-IR+UV)$ is lower than about
one third of SFR$(UV, corr)$, use only the 24$\mu$m based SFR
(SFR$(mid-IR)$) as the best estimate. These sources are not likely to
have the UV/optical light dominated by young massive stars, and are
probably quiescent or post starbursts (Fig.~\ref{fig:who_right}).

\item[3)] In the cases where SFR$(mid-IR+UV)$ is larger than about
3 times SFR$(UV, corr)$, i.e., for {\em mid-IR excess} galaxies, 
use SFR$(UV, corr)$ as the best SFR indicator, as it is likely
that
there is a contribution to the 24 $\mu$m flux density from other than star formation.
  Availability of deep radio imaging and/or submm/mm imaging,
can allow one to identify the subsample of galaxies among these that
are actually {\em opaque} to UV light, and longer wavelength estimates
of SFR are to be used in the latter.
\end{itemize}

To our understanding these recipes makes the best possible use of the
multiwavelength database currently available, in order to cope with
the limitations of each individual star formation indicator. In
particular, they are meant to single out those sources where a
particular method is bound to fail, e.g., because the UV slope
reflects age rather than reddening effects, or the mid-IR output is
partly powered by an AGN.

In principle, the similarity in the trends observed in Fig.~7--10 could 
suggest to define modified, highly non linear relations between 8\,$\mu$m 
luminosities and $L_{\rm IR}$ for $z=2$ galaxies, as suggested also by
Papovich et al. (2007), for recovering corrected $L_{\rm IR}$ for all
galaxies. However, Fig.~\ref{fig:corre} shows that in practice any such
relation would be dependent on the sample properties itself, e.g. shifting
as a function of $K$-band magnitude, and therefore not well defined
in general.

\section{Implications}
\label{sec:impli}

In this section, we exploit the results presented so far, and in
particular the improved SFRs for our sample of star-forming galaxies
at $z\sim2$, in order to address a number of issues which are relevant
for our understanding of the nature of star-formation and mass
assembly in such galaxies.

\subsection{The cosmic evolution of ULIRGs}

As already shown by D05b (see also Papovich et
al.\ 2006; Reddy et al.\ 2006a; Caputi et al.\ 2007), the typical massive
($M\simgt10^{11}M_\odot$~yr$^{-1}$) star-forming galaxy at $z=2$ is a
ULIRG, i.e., it appear to have $L_{\rm IR}\sim10^{12}L_\odot$ or more.  Here,
we provide a more detailed estimate of the space density of $z=2$
ULIRGs
using multiwavelength information.  In GOODS-N there 
are 113 $BzK$ galaxies that are classified as ULIRGs based on their 24
$\mu$m flux densities, 58 based on UV SFRs, and 45 based on radio emission (or up 
to 69 if considering the undetected radio sources with limits still
consistent with the ULIRG regime).  Similarly, in GOODS-S there are
155 $BzK$s classified as ULIRGs from their 24 $\mu$m flux densities, and 50 from
the UV SFRs.  At luminosities of $L_{\rm IR}=10^{12}L_\odot$ the mid-IR 
starts to be systematically in excess of radio (Fig.~\ref{fig:xlum_N}), and
therefore ULIRG densities derived from MIPS \24mu data are likely to
be overestimated. We notice however that the number of ULIRGs inferred from 
MIPS 24$\mu$m emission does not increase appreciably when moving from the
sample 
with $K<20.5$ in GOODS-N to the
one limited at $K<22.0$ in GOODS-S. On the other 
hand, there can be some sources that are not classified as ULIRGs in the UV
due to obscuration. Also, the radio flux density limits over this redshift range 
are very close to the ULIRG detection threshold, and hence a radio-based 
estimate could be somewhat incomplete. Therefore, we average 
the UV/radio and MIPS 
estimates to derive ULIRG space densities at $z\sim2$.  Combining GOODS-N (154
arcmin$^2$) and GOODS-S (140 arcmin$^2$ used in this work), and
correcting the ULIRG counts to compensate for the fact that we are
including only sources not blended in the IRAC images, we finally
derive a sky density of ULIRGs of 0.6 arcmin$^{-2}$. Using the
comoving volume within $1.4<z<2.5$ (Fig.~\ref{fig:zhist_N}), this
implies a space density of $1.6\times10^{-4}$~Mpc$^{-3}$.  The 
uncertainty in this density is order of 0.2~dex, when accounting for
the scatter in the ULIRG counts from the different $L_{\rm IR}$ tracers, as mentioned above. We note that we have excluded galaxies
with hard X-ray detections from this analysis as likely hosting
AGNs.  
If we use the 24$\mu$m emission
for the X-ray AGNs to estimate $L_{\rm IR}$, the sample of ULIRGs increases
by 15-20\%, thus still within the range of uncertainties estimated above. 
The space density of ULIRGs at $z=2$ is a factor of 1000 larger than in
the local Universe (e.g., Sanders er al.\ 2003), confirming the
previous estimate of D05b.

\begin{figure}    
\centering 
\includegraphics[width=8.8cm]{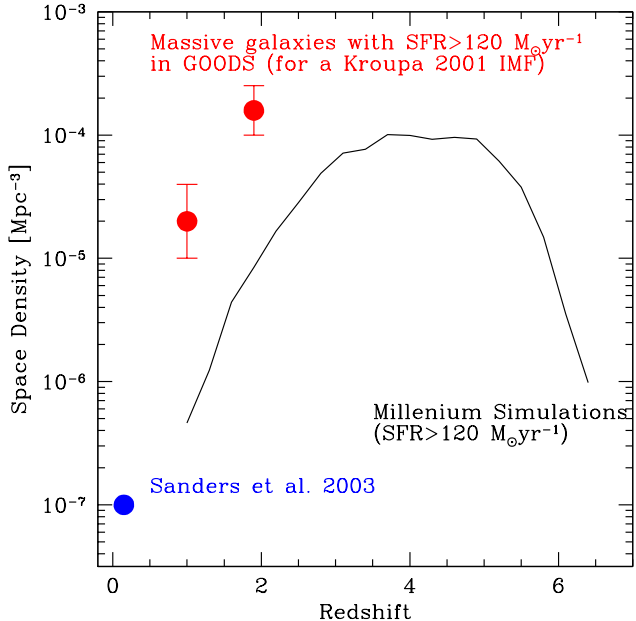}
\caption{The space density of ULIRGs (excluding individually detected AGNs) 
at $z=2$ and $z=1$ inferred from GOODS (red filled circles) is
compared to the local density from Sanders et al.\ (2003; filled green circle) and to predictions 
of star-forming galaxy density from the mock lightcones based on the Millennium simulation (Kitzbichler
\& White 2007). The rate of 120\,$M_\odot$~yr$^{-1}$ corresponds to a ULIRG in
the case of a Kroupa (2001) or Chabrier (2002) IMF.
}
\label{fig:ulirgs_Mille}
\end{figure}

In a similar way, we derive the space density of ULIRGs in GOODS at $z=1$, 
taking galaxies with $0.7<z<1.3$ (using photometric and spectroscopic redshifts)
and estimating $L_{\rm IR}$ using \24mu flux densities. We find only 7 such sources in the 
combined GOODS, which implies a space density of $2\times10^{-5}$, an order of 
magnitude lower than that at 
$z=2$.\footnote{Adding hard X-ray detected galaxies 
only marginally increases the estimated ULIRG space density
also at $z \approx 1$.}

In Fig.~\ref{fig:ulirgs_Mille} we compare these results to predictions from 
the Millennium simulation of galaxy formation (Springel et al.\ 2005).  
We use the lightcones based on semianalitic models presented by 
Kitzbichler \& White (2007). 
For a Kroupa et al.\ (2001) or Chabrier (2003) IMF
the ULIRG regime corresponds to a star-formation rate 
of 120\,$M_\odot$yr$^{-1}$, lower than the rate reported 
in Eq.~1 for a Salpeter IMF. We considered all simulated galaxies with SFR
above this threshold as ULIRGs. We explicitly quote here results for the
Kroupa (2001) and Chabrier (2003) IMFs
because these are favored by observations in the local Universe and therefore
more appropriate for comparison to predictions of SFRs in galaxies.
The model underpredicts the space density of ULIRGs at both $z=1$ 
and $z=2$ by at least an order of magnitude.
Only at higher redshifts, $3 < z < 5$, do the Millennium models predict a ULIRG space 
density as large as that which we observe at $z \approx 2$.  This suggests that star-formation 
in massive galaxies occurs too early in the simulations, or was terminated too early due to 
the adopted truncation of SFR from AGN feedback (e.g., Croton et al.\ 2006). 
However, we reach similar conclusion when comparing to simulated
galaxy catalogs based on the model of Oppenheimer \& Dav{\'e}  (2006), 
which also predicts a comparably large space density of ULIRGs only 
at $z \gg 2$.

\begin{figure}    
\centering 
\includegraphics[width=8.8cm]{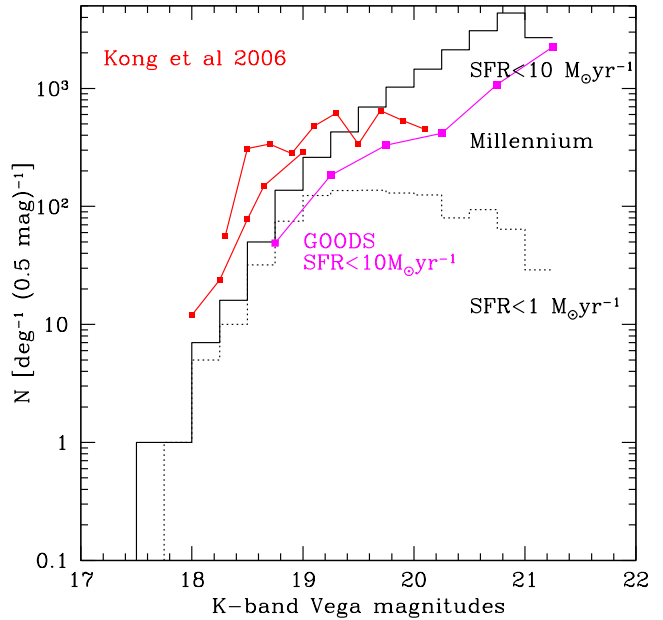}
\caption{The number counts of passive galaxies is plotted versus K-band magnitude.
The red points show the $pBzK$ counts from the two fields of the
large area survey of Kong et al.\ (2006).
Predictions obtained from the mock lightcones based on the Millennium simulation (Kitzbichler
\& White 2007) are also shown, for galaxies with $z>1.4$ and SFR\,$<1$ and 10 $M_\odot$yr$^{-1}$.
}
\label{fig:pBzK_Mille}
\end{figure}

\subsection{Counts of passive galaxies}

One possible means for reconciling the models with the observations would be to delay 
star-formation until later epochs.  Therefore, it is important to compare observations 
and model predictions for the abundance of passive, non-star-forming galaxies at $z \approx 2$.
In fact, the feedback from AGNs has been implemented in the Millennium models in order 
to allow them to reproduce the existence of passive and massive galaxies both at high and low 
redshifts (de Lucia et al.\ 2006). The Oppenheimer \& Dav{\'e} (2006) models, lacking the crucial mechanism
of AGN feedback, predict very low space 
density of passive galaxies at $z\sim$1--2, much below the observed density.

In Fig.~\ref{fig:pBzK_Mille} we show the number counts of {\em passive} $BzK$ ($pBzK$)
galaxies from Kong et al.\ (2006).
These objects, selected according to the criteria defined in D04b, are expected to provide
a fair census of the population of $z>1.4$ passively evolving systems (see, e.g., Daddi et al 2005a; Arnouts et al. 2007).
In GOODS the B-band data are not deep enough to reliably distinguish passive from 
star-forming $BzK$ galaxies using the color definition of D04b. However,
we may approximate this by using the number counts of GOODS $BzK$ galaxies that have 
SFR\,$<10\ M_\odot$yr$^{-1}$ based on the \24mu measurements.  As we have 
shown in Fig.~\ref{fig:pl_limi}, 
the MIPS \24mu observations 
are very sensitive for detecting low rates of star formation, 
and in particular to identify galaxies with red UV slopes but relatively low
SFRs (Fig.~\ref{fig:who_right}).
The counts of galaxies with  SFR\,$< 10\ M_\odot$yr$^{-1}$ from the combined GOODS N\&S fields
are shown in Fig.~\ref{fig:pBzK_Mille}. These are lower than the $pBzK$ counts of Kong et al.\ (2006),
most likely due to cosmic variance, as these sources are very strongly clustered (Kong et al.\ 2006).
The larger COSMOS survey may provide more robust statistics on the abundance of passive
galaxies, although we note that the \24mu    data in COSMOS (Sanders et al.\ 2007) are shallower
and cannot set such tight constraints on low SFRs at $z \approx 2$.

Overall, the Millennium simulations predictions, shown in Fig.~\ref{fig:pBzK_Mille},
seem to reproduce the data reasonably well, given the large uncertainties, 
perhaps somewhat overestimating the counts of galaxies with relatively 
low SFRs. 

\subsection{The duty cycle of $z\sim2$ ULIRGs}

The Millennium models might under-predict the ULIRG space density if the star 
formation episodes were short lived, so that the peak SFR is missed caught when 
averaging over the simulation time steps (250~Myr at $z=2$ for the evolution of 
the dark matter, with much finer steps for physical processes computed in the
semi-analytic models).  It is therefore important to try to constrain the duration 
of the phase at high redshifts. This is also relevant in order to evaluate the 
contribution from this star-formation mode to the assembly of stars in massive 
galaxies. In D05b it was suggested that duty cycles are high, 
based on the very large detection rate and brightness of the most massive galaxies 
at \24mu. Caputi et al.\ (2006) agree with this conclusion, based on their \24mu 
study of $z=2$ galaxies in GOODS.  We investigate this point in more detail here, 
in the light of our new results on the comparison of the different star formation 
estimators.

\begin{figure*}    
\centering 
\includegraphics[width=18cm]{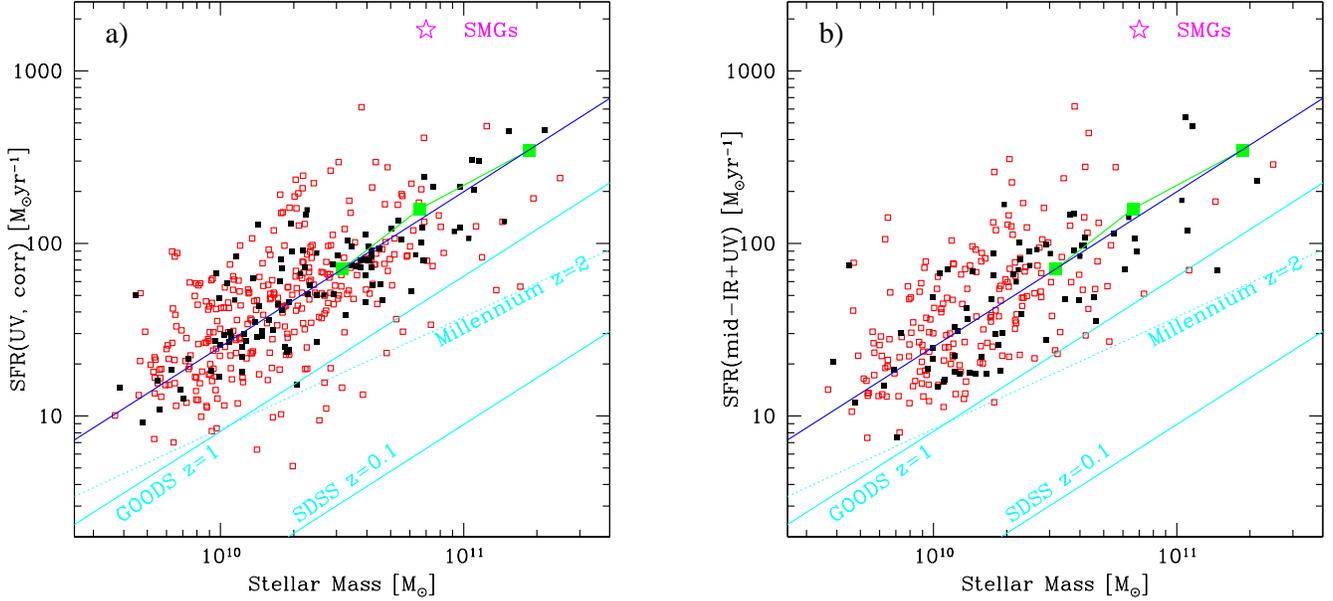}
\caption{The Stellar Mass-SFR correlation for $z=2$ star forming galaxies in GOODS. Points are taken from 
the deeper GOODS-S field to $K=22$. We include only \24mu detected galaxies:
passive/quiescent galaxies are excluded from this analysis.
Left panel (a) has SFRs derived from UV, corrected for dust extinction.
Right panel (b) has \24mu derived SFRs but we explicitly exclude all
mid-IR excess galaxies.
The large green dots are the result of average SFR-Mass relation in GOODS-N
determined from radio stacking of $K<20.5$  galaxies in 3 mass bins. The blue line is 
SFR\,$=200\times M_{11}^{0.9}$ $[M_\odot$ yr$^{-1}]$, where $M_{11}$ is the stellar mass in units
of $10^{11}M_\odot$. The cyan solid lines are the $z=1$ and $z=0$ correlations, taken from Elbaz et al.\ (2007),
that have a similar slope of 0.9.
The cyan dashed line is a prediction for $z=2$ from the Millennium simulations, based on the mock
lightcones of Kitzbichler \& White (2007). The magenta star near the top shows the location of typical
SMGs in this diagram.
}
\label{fig:SFR-Mass}
\end{figure*}

In order to do this, we consider a mass-- and volume-limited sample of $BzK$ galaxies,
and evaluate the fraction of ULIRGs in the sample. If ULIRGs stopped forming stars, 
they would remain in the mass-selected sample as passive or slowly star-forming 
galaxies.  The fraction of ULIRGs in this mass-selected sample is therefore a lower 
limit to their duty cycle, as the ULIRG phase could have started well before the
galaxies reached the required mass threshold, or at an earlier era, before the redshift 
boundary of the sample considered here.

To build a mass-limited sample of $z\sim2$ galaxies, we consider all of the 
$z>1.4$ $BzK$ galaxies, including both star forming and passive galaxies.  
The latter were excluded for the analyses in previous parts of the paper.
We use the empirical mass estimate described in D04b that is 
based on detailed SED fitting results with the Bruzual \& Charlot (2003) models 
performed by Fontana et al.\ (2004).  This calibration allows us to associate 
a mass to each galaxy based on its K-band magnitude, with a correction for 
varying $M/L$ ratio based on the observed $z-K$ color.  
Maraston et al.\ (2006; 2007) have shown that for passive galaxies, where 
the overall SED is dominated by $\approx0.5$--1~Gyr old stars, the use of the 
Maraston (2005) model libraries results in masses that
are smaller by 0.2--0.3~dex (see also Bruzual 2007).  
For star-forming $BzK$ galaxies, where the young stellar 
component dominates the light, we find (Maraston et al., in preparation) that the 
Maraston models provide $M/L$ ratios that are consistent on average with those from
Bruzual \& Charlot (2003).  

In the $M > 10^{11} M_\odot$ sample constructed in this manner, we find 34 galaxies at 
$z\sim2$ in GOODS-S and $38$ in GOODS-N. Only 1/34 of the GOODS-S galaxies above this 
mass threshold has $K>20.5$, which suggests that the sample is fairly complete also 
in the GOODS-N to this limit. The space density of massive galaxies with
$M>10^{11}M_\odot$ is about $8\times10^{-5}$~Mpc$^{-3}$. This estimate is within
a factor of $\approx2$ of the one reported by van Dokkum et al. (2006) using the
larger MUSIC survey, despite the different methods for estimating stellar 
masses. 

In order to estimate SFRs, we use both MIPS \24mu and the UV luminosities corrected for reddening.
While we have shown that SFRs inferred from \24mu are suspect if they are substantially larger those
derived from the UV data, the MIPS data are also very efficient for measuring relatively low SFRs in 
red galaxies where UV-reddening estimates would be in turn overestimated (see Fig.~\ref{fig:who_right}). 
Therefore, the MIPS data are ideal for isolating the most quiescent galaxies even in the case that 
some low level residual SFRs place these sources in the star forming $BzK$ region.

By taking the minimum SFR for each object, estimated from either the \24mu 
flux density or from the 
extinction--corrected UV luminosity, we find that the fraction of ULIRGs among the 
most massive galaxies at $z \approx 2$ is $39\pm5$\% (Poisson counting error only). 
This very conservative estimate sets a similar constraint on the duty cycle of the ULIRG 
phase in massive high-$z$ galaxies, i.e., $\simgt$40\%. The cosmic time span elapsing 
during the redshift range $1.4<z<2.5$ is about 2~Gyr, with only 1~Gyr per galaxy 
available on average (given the reasonably flat redshift distribution within these limits).
This implies that these ULIRG phase will last for at least some 400~Myr, one order of 
magnitude longer than what is currently estimated for SMGs at similar redshifts, or 
for local ULIRGs (see, e.g., Greve et al.\ 2005; Tacconi et al.\ 2006; although 
see also Swinbank et al.\ 2006).

\begin{figure}    
\centering 
\includegraphics[width=8.8cm]{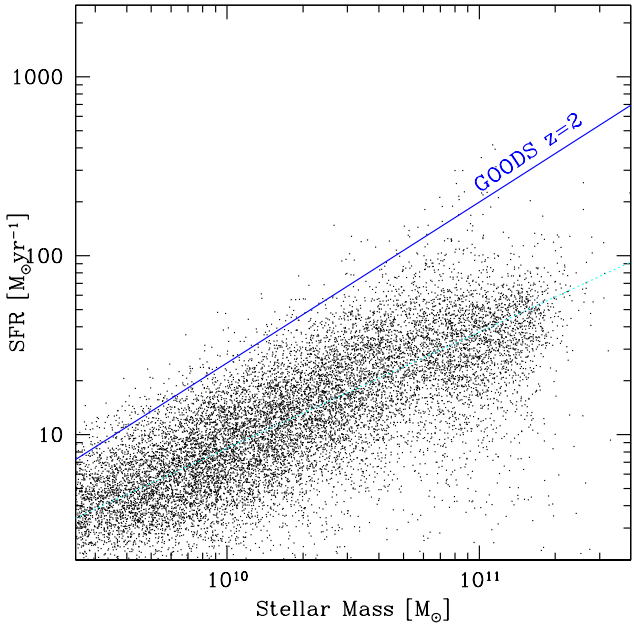}
\caption{The same of Fig.~\ref{fig:SFR-Mass}, but here simulated galaxies at $z=1.90\pm0.05$ are plotted,
taken from the Millennium lightcones of Kitzbichler \& White (2007). The cyan line is a power law with 
slope of 0.65, and the blue solid line is the same 
of the one in Fig.~\ref{fig:SFR-Mass}, showing the average
trend for the GOODS $BzK$ sample at the same redshift.
}
\label{fig:SFR-Mass-Mille}
\end{figure}
\subsection{The star formation mass correlation at $z=2$}

We can gain further insights on the nature of star-formation at $z=2$ by examining 
correlations between galaxy mass and star formation rates.  Recently, Noeske et al.\ (2007) 
and Elbaz et al.\ (2007) have shown that star formation and stellar mass define
a tight correlation (0.2~dex scatter) in galaxies at $z \sim 1$, 
with rough proportionality 
(logarithmic slope of 0.9). 
Similar proportionality is also seen at $z=0$ in data from
the Sloan Digital Sky Survey (Elbaz et al.\ 2007), although with a lower normalization 
reflecting the overall decline in cosmic SFR density with time.

Fig.~\ref{fig:SFR-Mass}
(left panel) shows the SFR- stellar mass relation for galaxies at $z \approx 2$ 
in GOODS-S using the UV to derive SFRs. In this plot, we show only those sources with 
MIPS detection,  with \24mu-derived SFR not less than 1/2 of the UV one (to avoid 
estimating SFR from UV for evolved galaxies, see Fig.~\ref{fig:who_right}) and with
a well-defined UV slope for obtaining an accurate reddening estimate. 
We find that a proportionality with a logarithmic slope of 0.9 (blue line in Fig.~\ref{fig:SFR-Mass}), 
normalized to the median SFR/M$_*$ ratio of the sample (2.4 Gyr$^{-1}$), provides a good fit to the
data, with semi interquartile range of only 0.16 dex in the SFR/M ratio. 
We have cross-checked the
validity of this relation in GOODS-N using the deep radio observations. 
We have divided the galaxies
(limited to $K=20.5$, a smaller range than what plotted in Fig.~\ref{fig:SFR-Mass}) in 3 independent 
mass ranges and obtained the average SFR in the bin using the radio, 
averaging together radio detections and non detections.   The green points in Fig.~\ref{fig:SFR-Mass} 
show that the star formation results from the stacked radio emission agree very well with the
relation that we derive from the UV light.  This check is important because it is based on an indicator 
that should, in principle, be more solid than UV, and also because it allows us to include many of 
the massive galaxies for which UV SFR could not be estimated due to their faintness in the optical 
bands.

Due to the substantial presence of mid-IR excess objects at $z \approx 2$, we do not expect that this 
relation can be accurately recovered using the \24mu inferred SFRs.  This explains why Caputi et al.\ (2006b) 
find a much looser correlation between SFR and masses for $z=2$ galaxies.  If we use \24mu inferred SFR only 
for those galaxies where this quantity is within a factor of 
3 of the UV one (the mid-IR normal 
objects), we obtain a SFR-mass correlation fully consistent 
with that based on UV 
(Fig.~\ref{fig:SFR-Mass}, right panel). 

The $z=2$ correlation appears to have a slope similar to that at lower redshifts.  Instead, the 
normalization at $z=2$ is a factor of 3.7 larger than that at $z=1$, and 27 times larger than at 
$z=0$ (Elbaz et al.\ 2007; Noeske et al.\ 2007).  At fixed stellar mass, star forming galaxies 
were much more active on average in the past.  This is most likely due to a larger abundance 
of gas, depleted with passing of time. 

The inferred correlation is quite tight, with semi interquartile range of only 0.16 dex in the
dispersion of specific SFRs.  We caution that, having used
mainly the UV as a SFR estimator, we cannot 
reliably rule out the presence of a larger number of outliers at 
low stellar masses, for which we might strongly 
underestimate the SFR from the UV.  These cannot be reliably identified to meaningful depths 
with radio data (due to the flux density limits of current observations), nor at \24mu (due to the 
existence of mid-IR excess sources).   We will have to wait for Herschel and ALMA to address 
this issue accurately. 

Submillimeter-selected galaxies are strong outliers to this trend, however.  
Tacconi et al.\ (2006) estimate that SMGs
in their sample have typically $L_{\rm IR}=10^{13}L_\odot$ and dynamical masses $\simlt 10^{11}M_\odot$ (see also
Greve et al.\ 2005 for similar results). For a given stellar mass (assuming that most of the dynamical mass
in the central regions of SMGs is stellar), SMGs are forming stars at a 10x or larger rate
respect to ordinary massive star forming galaxies. Their space density is also approximately an order of 
magnitude smaller.
SMGs at $z=2$ appear to be like LIRGs and ULIRGs at $z=0$, i.e., relatively rare objects and outliers 
of the mass-SFR correlation (see
Elbaz et al.\ 2007). To this regard, it is not surprising that much shorter star formation duty cycles 
and lifetimes have been inferred for SMGs and local ULIRGs ($\ll 100$Myr; Greve et al.\ 2005; 
Solomon \& Vanden Bout 2005).  These represent short-lived stages of the life of galaxies, 
due, e.g., to ongoing mergers or some temporary
perturbations (see, e.g., Dannerbauer et al.\ 2006 for a discussion). 
ULIRGs and SMGs also generally have smaller physical sizes (Tacconi et al.\ 2006) than
those of more ordinary massive, star-forming galaxies at $z = 2$ 
(D04a; Ravindranath et al.\ 2007, in preparation), suggesting that the SMGs
are in more advanced merger states.

We have used the mock lightcones from  Kitzbichler et al.\ (2007), based on the Millennium simulations, 
to explore the comparison of mass and SFR at $z=2$ in these models (Fig.~\ref{fig:SFR-Mass-Mille}). As emphasized already by Finlator et al. (2006)
theorethical simulations quite naturally predict 
the existence of correlations 
between galaxies SFRs and stellar masses. However, 
we find that at fixed stellar masses, the model
galaxies are forming stars at about 1/4 of the observed rate for galaxies with $M\sim10^{11}M_\odot$. 
The correlation is also substantially tilted,
with decreasing specific SFR at larger masses. 
It seems that a major change required for models would be to 
increase the star-formation efficiency (and thus the typical SFR) at all masses for star-forming galaxies
at  redshifts $0.8<z<3$, while still keeping  the current proportion of massive galaxies in a 
passive/quiescent state.  Interestingly, we find that if we reproduce Fig.~\ref{fig:SFR-Mass-Mille}
plotting simulated Millennium galaxies at $z=3$ instead of $z=1.9$, we find that simulated galaxies match 
the $z=2$ GOODS galaxies quite accurately, 
with the same SFR versus mass normalization, slope and with 
similar scatter. This again reinforces the idea that SFR and mass 
growth happens too early in the current version of the simulations. 

The observed mass-SFR correlation defines a more basic 
dichotomy of galaxy properties than those based
on colors, as emphasized also by Elbaz et al.\ (2007). 
In future papers, we will investigate other physical properties 
of $z=2$ galaxies (like morphology)
as a function of their distance from the correlation.
Finally, we note that the trend of SFR increasing with galaxy mass, 
and the fact that star 
formation is sustained with long duty cycles, formally imply very rapid 
growth of the stellar mass during 
the period spanned by the observations.
We confirm the conclusions of D05b, that the epoch within 
$1.4<z<2.5$ corresponds to the major build-up period of the most massive 
galaxies in the local Universe.

\section{Summary and conclusions}
\label{sec:end}

We have investigated the properties of K-selected, star forming galaxies at $z \sim 2$ in the GOODS fields,
by means of over 1000 galaxies with $1.4\simlt z \simlt 2.5$, several hundred of which have measured
spectroscopic redshifts.  We have used the deepest observations at a variety of wavelengths, including
radio, Spitzer \24mu and \70mu, SCUBA 850\,$\mu$m and UV light, in order to derive and compare
multiwavelength luminosities of galaxies, and to investigate the reliability of these luminosities for measuring 
star formation rates. Our findings can be summarized as follow:

\begin{itemize}

\item For most galaxies with moderate mid-IR luminosities, $L(8\mu m) < 10^{11}L_\odot$, 
Spitzer \24mu measurements can be used to estimate total infrared luminosities and star formation
rates of galaxies at $z\sim2$.  The ratios of the mid-IR luminosities to those at other wavelengths 
(radio, \70mu, 850\,$\mu$m and even reddening-corrected UV light) are consistent with expectations from 
local correlations, provided that their mid-IR luminosities is not too large.  

\item However, most galaxies with $L(8\mu m)>(1$--$2)\times10^{11}L_\odot$ show a mid-IR excess  
with respect to all other luminosities, compared to that expected from local correlations, and on ratios 
observed in less luminous $z=2$ galaxies.  We therefore report the discovery of a large population of 
$z\sim2$ galaxies with intrinsic mid-IR excess.  We investigate the nature of these sources in a 
companion paper (Daddi et al.\ 2007; \hbox{Paper~II} hereafter), 
concluding that these objects contain heavily obscured, often 
Compton thick, AGN. 

\item The presence of galaxies with seemingly extreme $SFR \gg 1000$ $M_\odot$~yr$^{-1}$ estimated from \24mu    
flux densities is nearly always due to the existence of these mid-IR excess sources.   A large proportion of \24mu 
sources at $z \sim 2$ detected in Spitzer surveys shallower than GOODS will be mid-IR excess objects,
and their star formation rates and infrared luminosities may be greatly overestimated if bolometric 
corrections from local starburst galaxy templates are applied.  The true SFR of these objects rarely 
exceeds a few hundred $M_\odot$~yr$^{-1}$.   

\item For typical $z=2$ galaxies, observed at fainter 24$\mu$m flux densities, 
the local correlations among the
SFR tracers that we have probed here still hold within the uncertainties.

\item The observed UV luminosity, corrected for dust reddening based on the observed UV slope using a Calzetti
et al.\ (2000) extinction law, allows one to obtain reliable estimate of the SFR activity for a large majority
of $z=2$ galaxies. This is demonstrated in comparison with mid-IR estimates (except for sources with intrinsic
mid-IR excess), and with radio, \70mu and 850\,$\mu$m observations through stacking in bins of UV and mid-IR 
luminosity.  The typical ULIRG at $z=2$ is 'transparent' to UV light, contrary to present--day ULIRGs and to 
$z=2$ SMGs.

\item As exceptions, we do find a minority of $z\sim2$ galaxies which appear to be opaque to UV emission
from star formation.  Even after correction for dust reddening, the SFR derived from the UV light in
these objects is underestimated by large amounts.  These objects are similar to SMGs, and 
given the depths of current GOODS multiwavelength data, they are best recognized as highly obscured 
outliers by comparing their UV and radio properties.

\item Similarly, there is a small population of K-selected sources that are classified as star forming
galaxies based on optical-UV colors (e.g., with the $BzK$ diagram) but for which the UV estimates of SFR greatly
overpredict the true amount of SFR. 
These are probably post-starburst galaxies, whose red UV 
colors are not only due to dust reddening, but also have important 
contributions from intrinsically cooler, lower-mass stars.  Also these objects 
can be identified from the comparison of UV and mid-IR based SFRs.

\item Exploiting all available $z=2$ SFR estimators, we have derived a measure
of the space density of ULIRGs at $z\sim2$ in the GOODS fields.  
We find a space density of about 
$2\times10^{-4}$~Mpc$^{-3}$ and a sky density of 0.6 arcmin$^{-2}$ 
at $1.4 < z < 2.5$.  This is a factor 
of 10 larger than the abundance predicted by semianalitic models based 
on the Millennium simulations.

\item Similarly, we have used our improved understanding of $z=2$ SFR tracers to investigate the abundance
of $z\sim2$ massive galaxies with low ongoing SFR (below 10~$M_\odot$~yr$^{-1}$). We have computed number
counts of these galaxies as a function of their K-band magnitude, and find that the counts are lower than 
that of passive ($pBzK$) objects selected by Kong et al.\ (2006), likely due to cosmic variance and possibly to 
some contamination of $pBzK$ samples from $z<1.4$ dusty galaxies. 
Predictions based on the Millennium simulations
can reasonably well account for the space density of these galaxies.\\

\item We have built a mass-limited sample of galaxies with $M>10^{11}M_\odot$ at $z\sim2$.
By estimating the distribution of SFRs in these galaxies, we derive lower limits to the duty 
cycle of ULIRGs. This is defined as the fraction of ULIRGs inside the mass-limited sample. We conclude
that this duty cycle is at least 0.4. This is much larger than the duty cycle estimated for local ULIRGs,
and also for $z=2$ SMGs. The relatively large duty cycles imply SFR duration at 
ULIRG levels of order of 0.5~Gyr or more. This also implies that large gas reservoirs have to exist 
in most of the massive $z=2$ galaxies, or that gas must be accreted very efficiently over time.

\item We have found that a relatively tight stellar mass-SFR relation is 
already in place at $z=2$, 
for star forming galaxies detected at \24mu, with a scatter 
(semi-interquartile range) of only 0.16~dex
in the specific star formation rate.  This has been derived mainly basing on 
the UV estimates of SFRs, and
confirmed through radio stacking of galaxies in mass bins and using SFRs derived
from the \24mu flux density, once mid-IR excess objects are excluded from the sample. 
The typical star forming galaxy
at $z=2$ is forming stars more rapidly by factors of 3.7 and 27 respect 
to an object with similar mass
at $z=1$ and $z=0$.   Current numerical simulations are able to reproduce 
the mass-SFR correlation with similar 
slope and similarly small scatter to what observed, 
but with much lower normalization than 
the observed one.

\item We conclude by suggesting that current realizations of galaxy 
formation models could come to fairly
good agreement with the observations available for $z=2$ galaxies if, by some means, the major periods of
star formation activity could be delayed until later epochs than in current implementations.\\

\end{itemize}

\acknowledgements
We thank the rest of the GMASS team for allowing us to use
the still unpublished spectroscopic redshifts, and the the many
other members of the GOODS team, 
who have helped to make these observations possible.
We are grateful to Emily MacDonald, Daniel Stern and Hy Spinrad
for collecting some of the redshifts
used in this work, and to Manfred Kitzbichler, Romeel Dav{\'e} and 
collaborators for the galaxy catalogs based on their model predictions. 
ED gratefully acknowledges NASA support (at the beginning of this work)
through the Spitzer 
Fellowship Program, award 1268429.
DMA thanks the Royal Society for funding. 
JK acknowledges financial support from the German Science Foundation (DFG)
under contract SFB-439.
Support for this work, part of the Spitzer Space Telescope
Legacy Science Program, was provided by NASA, Contract Number 1224666
issued by the JPL, Caltech, under NASA contract 1407.

\end{document}